\documentclass[
	aps,
	prl,
	amsmath,
	reprint,
	superscriptaddress,
	longbibliography
]{revtex4-1}
\usepackage{hyperref}
\usepackage{graphicx}
\usepackage{units}
\usepackage{upgreek}
\usepackage{color}
\usepackage{amsmath, amssymb, amsfonts}
\makeatletter
\renewcommand\subparagraph{\@startsection{subparagraph}{5}{\z@}{3.25ex\@plus1ex\@minus.2ex}{-1em}{\normalfont\normalsize\bfseries}}

\renewcommand{\@make@capt@title}[2]{\@ifx@empty\float@link{\@firstofone}{\expandafter\href\expandafter{\float@link}}{\textsf{\bfseries#1}}\@caption@fignum@sep{\textsf{#2}}}

\renewcommand\@caption@fignum@sep{\textsf{\bfseries.}}

\let\oldtheequation\theequation
\renewcommand\tagform@[1]{\maketag@@@{\ignorespaces#1\unskip\@@italiccorr}}
\renewcommand\theequation{(\oldtheequation)}
\makeatother

\usepackage{tabularx}

\begin{document}

\begin{abstract}
	In an equilibrium thermal environment, random elastic collisions between background particles and a tracer establish the picture of Brownian motion fulfilling the celebrated Einstein relation between diffusivity and mobility. In nature, environments often comprise collections of autonomously moving objects that exhibit fascinating non-equilibrium phenomena and are nowadays termed ``active matter''. We investigate experimentally the impact of an active background on a passive tracer using vibrationally excited active particles. They produce multiple correlated ``tapping collisions'' with the tracer, for which a persistent memory emerges in the dynamics. The system is described by a generalized ``active'' Einstein relation that constrains fluctuations, dissipation, and effective activity, by taking the emerging tracer memory into account. Since the resulting persistence can largely be tuned by the environmental density and motility, our findings can be useful to engineer properties of various active systems in biomedical applications, microfluidics, chemical engineering, or swarm robotics.
\end{abstract}

\title{Emergent memory from tapping collisions in active granular matter}

\author{Lorenzo Caprini}
\email{lorenzo.caprini@gssi.it} 
\affiliation{Institut f\"ur Theoretische Physik II: Soft Matter, Heinrich-Heine-Universit\"at D\"usseldorf, D-40225 D\"usseldorf, Germany}

\author{Anton Ldov}
\affiliation{Institut f\"ur Theoretische Physik II: Soft Matter, Heinrich-Heine-Universit\"at D\"usseldorf, D-40225 D\"usseldorf, Germany}

\author{Rahul Kumar Gupta}
\affiliation{Institut f\"ur Theoretische Physik II: Soft Matter, Heinrich-Heine-Universit\"at D\"usseldorf, D-40225 D\"usseldorf, Germany}

\author{Hendrik Ellenberg}
\affiliation{Institut f\"ur Theoretische Physik II: Soft Matter, Heinrich-Heine-Universit\"at D\"usseldorf, D-40225 D\"usseldorf, Germany}

\author{Ren\'e Wittmann}
\affiliation{Institut f\"ur Theoretische Physik II: Soft Matter, Heinrich-Heine-Universit\"at D\"usseldorf, D-40225 D\"usseldorf, Germany}

\author{Hartmut L\"owen}
\affiliation{Institut f\"ur Theoretische Physik II: Soft Matter, Heinrich-Heine-Universit\"at D\"usseldorf, D-40225 D\"usseldorf, Germany}

\author{Christian Scholz}
\affiliation{Institut f\"ur Theoretische Physik II: Soft Matter, Heinrich-Heine-Universit\"at D\"usseldorf, D-40225 D\"usseldorf, Germany}

\date{\today}

\maketitle

\section{Introduction}
Active matter comprises a new material class of agents that self-propel by converting energy from their environment into directed motion~\cite{marchetti2013hydrodynamics, Elgeti2015, bechinger2016active}. Important examples are living systems, such as flocks of birds~\cite{mora2016local}, schools of fish~\cite{ward2008quorum}, and swarming of bacteria~\cite{peruani2012collective, wioland2016ferromagnetic, liu2021viscoelastic}, or artificial particles, like drones~\cite{vasarhelyi2018optimized}, self-propelling robots~\cite{rubenstein2014programmable, Patterson2017, PhysRevX.6.011008}, active granulates~\cite{aranson2007swirling, kumar2014flocking, kudrolli2010concentration, baconnier2022selective}, spinners \cite{VanZuiden2016, scholz2021surfactants, lopez2022chirality}, and colloidal microswimmers~\cite{buttinoni2013dynamical, bricard2013emergence}. Due to their autonomous motion, active particles represent a system far from equilibrium~\cite{fodor2021irreversibility, o2022time} and behave fundamentally differently from equilibrium systems like thermal gases or liquids formed by passive particles.\par
\begin{figure}
	\includegraphics[width=\columnwidth]{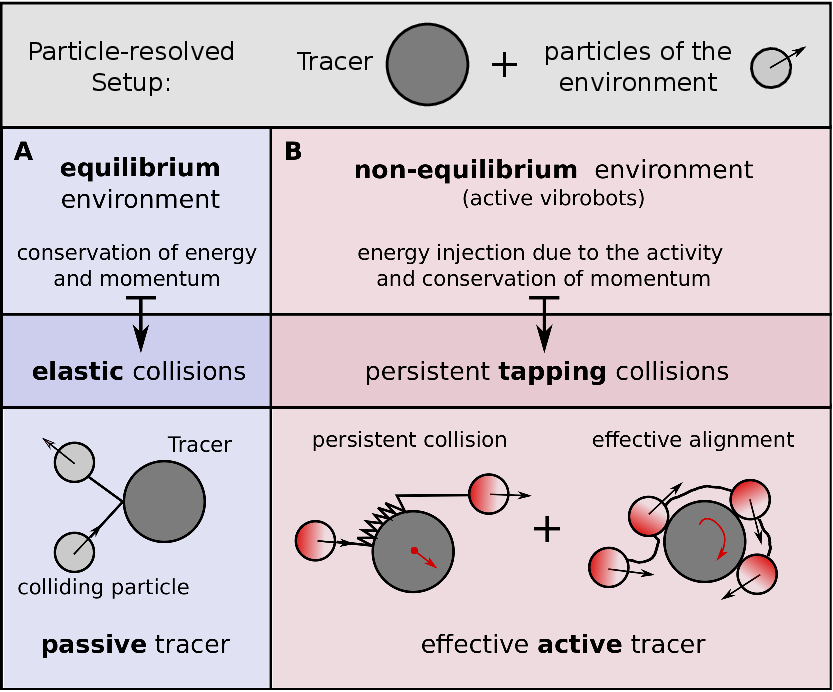}
	\caption{
		\label{Fig:Fig0}
		\textbf{Tracer dynamics in a (non-)equilibrium environment.}
		Tracer and bath particles are represented by large and small disks, respectively.
		(\textbf{A}) the equilibrium scenario is characterized by elastic collisions that conserve energy and momentum. After a collision with the environmental particle, the tracer behaves as a passive unit subject to Brownian motion and dissipation.
		(\textbf{B}) the non-equilibrium environment formed by active vibrobots does not conserve energy, which is continuously injected into the system by the active force. Particles of the environment display tapping collisions, i.e.\ collisions are persistent and the particles of the environment align around the tracer. As a result, the tracer performs an effective active motion. Note that, in general, collisions of granular particles are partially inelastic. However, this does not alter the present picture.
	}
\end{figure}

In thermal systems, when a tracer is surrounded by an environment of particles, it is kicked around by collisions, constantly exchanging momentum and energy with said environment. In a thermal equilibrium background, the collisions are elastic (\autoref{Fig:Fig0}~A) and the tracer is hit stochastically from all directions. Typically, for tracer sizes larger than those of the environmental particles, the resulting dynamics of the tracer is random Brownian motion~\cite{frey2005brownian, hanggi2005introduction} as established by  
Smoluchowski and Einstein more than a century ago~\cite{einstein1905motion}. 
%~\cite{von1906kinetischen} 
In this case, the tracer moves with negligible memory duration, completely forgetting the direction of the last kick by the time it experiences a new one. In other words, the directions of any two sequential collisions are not correlated.
Inertia is negligible in these systems, unless ultra-small timescales are resolved~\cite{kheifets2014observation}, and does not qualitatively affect the collision behaviour in equilibrium.

%sutherland1905lxxv, 

In an inertial active environment the situation is fundamentally different: Active particles hit the tracer and bounce back at first by momentum conservation. Soon, however, they recover due to their self-propulsion, and may return to kick the tracer again. This sequence of events can occur multiple times and sequential collision directions are thus highly correlated.
Despite this intuitive mechanism, systematic experimental studies on particle-resolved collisions in inertial active backgrounds are still unexplored. Their effects on the thermodynamic description of a tracer are therefore unknown.\par
To describe equilibrium systems, Einstein proposed and derived a famous fundamental relation, known sometimes as the Einstein-Sutherland-Smoluchowski relation~\cite{einstein1905motion, hanggi2005introduction, marconi2008fluctuation}, that links a system's fluctuation and dissipation: The product of mobility (dissipation) and system temperature (fluctuations) equals the tracer's diffusivity. The validity of the
Einstein relation has been confirmed for passive colloids~\cite{blickle2007einstein}, while its violation has been experimentally observed in systems intrinsically out of equilibrium~\cite{marconi2008fluctuation}, such as granulates~\cite{d2003observing, umbanhowar1996localized, RN5681, Agrawal2020}, glasses~\cite{crisanti2003violation}, and bacteria~\cite{maggi2017memory, di2010bacterial, sokolov2010swimming}. 
Intuitively, the slow relaxation of glasses or the extra injection of energy due to a non-equilibrium mechanism for granular particles and bacteria breaks the Einstein relation by increasing the fluctuations and changing the system's dissipative properties.
In general, this violation can be due to memory effects induced by mechanisms, such as non-reciprocity~\cite{loos2020irreversibility}, interactions with chemical trails~\cite{hokmabad2022chemotactic}, viscoelasticity~\cite{ginot2022barrier}, and particle collisions~\cite{thuroff2013critical}.
There exists in the literature no general theory for systems that are intrinsically out of equilibrium but, in active systems, such a violation has been recently interpreted as an effective temperature~\cite{petrelli2020effective, flenner2020active}.
However, extensions of the Einstein relation, linking dissipation, fluctuations, and non-equilibrium dynamical mechanisms, constitute an active field of research~\cite{marconi2008fluctuation, harada2005equality, verley2011modified, dinis2012fluctuation, neri2019integral, caprini2021generalized, solon2022einstein}.
\par

Here, we investigate the impact of an active environment on a passive tracer on the fundamental particle-resolved level, by observing the paths of a tracer in a bath of vibrationally excited active granular particles~\cite{deseigne2010collective, Koumakis2016}, so-called \emph{vibrobots}. Resolving the collisions between the tracer and active particles, we experimentally observe that the combination of bounce-back effects and activity results in multiple collisions of the same active particle on the tracer with an additional alignment mechanism. We refer to these multiple collision events as \emph{tapping collisions} (\autoref{Fig:Fig0}~B). These tapping collisions are a mark of the non-equilibrium nature of the bath and have a profound influence on the tracer dynamics.
In fact, they generate memory in the tracer motion resulting in a breakdown of the conventional Einstein relation.
Here, we propose a generalized active Einstein relation by incorporating the emergent activity explicitly, and present confirmation of our theory by particle-resolved experimental data.
We remark that a similar Einstein relation has been investigated numerically in Ref.~\cite{solon2022einstein}, but explaining our experimental results requires taking into account additional rotational dynamics that characterize our vibrobots. This induces an effective, increased persistence time and an Einstein relation for the rotational motion.
\par
We demonstrate that the memory time and the persistence length of the tracer can be efficiently tuned by the properties of the active environment, such as density and motility. This has two important consequences: Firstly, our findings can be useful to engineer tracer dynamics in various active backgrounds, with potential, future applications for two-dimensional swarm robotics.
%important for such diverse fields as drug delivery, microfluidics, or swarm robotics. 
Secondly, and more fundamentally, we can tune a tracer with a prescribed memory almost at will by adjusting the environment accordingly. This provides a realization of fundamental stochastic models for self-propelled particle motion~\cite{szamel2014self, maggi2015multidimensional, fodor2016far, caprini2019activityinduced, PhysRevLett.129.048002}.\par

\section{Results}\label{sec2}

\begin{figure*}
	\includegraphics[width=\textwidth]{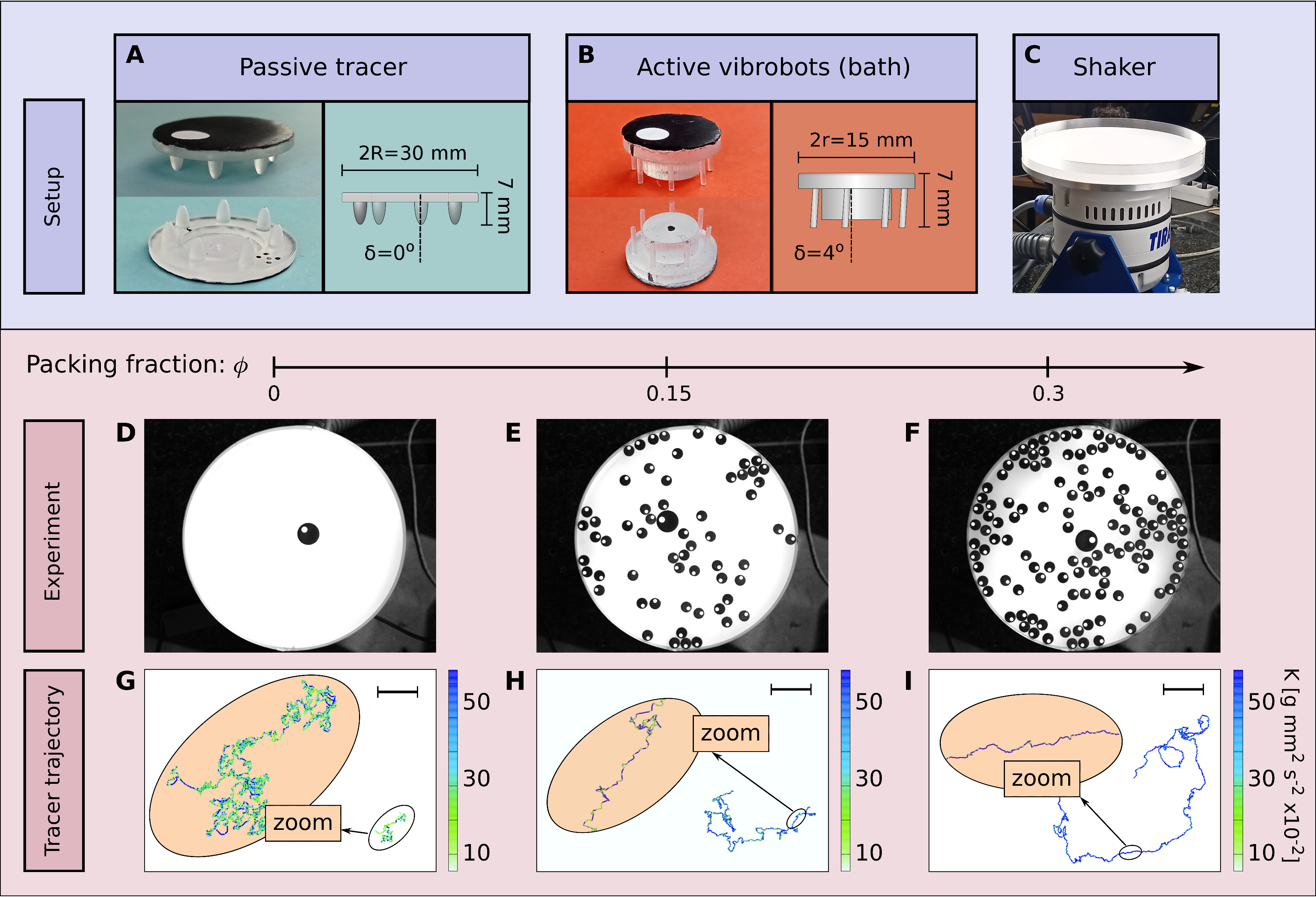}
	\caption{
		\label{Fig:Fig1}
		\textbf{Experimental setup: Passive tracer in an inertial active bath.}
		(\textbf{A}-\textbf{B}) Photos and diagrams of passive tracer and active vibrobots, respectively. White spots indicate a reference angle for the passive tracer and the orientational angle for active particles.
		(\textbf{C}) Photo of the shaker and plate where the motion takes place.
		(\textbf{D}-\textbf{F}): Captured images of the passive tracer in a bath of active particles at packing fractions $\phi=0,0.15,0.3$, respectively.
		Corresponding $\sim\unit[80]{s}$ long tracer trajectories are shown in (\textbf{G}-\textbf{I}), with a color gradient denoting the total kinetic energy of the particle $\mathcal{K}=\mathcal{K}_\text{t}+\mathcal{K}_\text{r}$, given by the sum of the translational and rotational contributions. Trajectories show an emergent memory for the active bath that results in correlated motion.
		Here, the scale bars read $\unit[55]{mm}$.
	}
\end{figure*}

\subparagraph{Experimental system.}
We experimentally study the dynamics of a passive tracer in a bath of inertial active particles.
Passive and active particles are 3D-printed vibrobots (see \autoref{Fig:Fig1}~A-B) driven by sinusoidal vibrations provided by an electromagnetic shaker to a circular plate, which the particles are free to explore (see \autoref{Fig:Fig1}~C).
Under these conditions, particles perform quasi-two-dimensional dynamics since their motion along the vertical direction is not relevant. What differentiates the tracer from the active bath particles are their respective shapes, described in detail in the Methods section, along with resultant physical properties. An active particle's tilted elastic legs (\autoref{Fig:Fig1}~B) provide an asymmetry in the direction of relaxation, thus converting energy, harvested from impacts with the vibrating plate and stored as elastic deformation, into directed motion. At long times, its motion is randomized due to surface inhomogeneities and bouncing instabilities that produce small random reorientations of the particle~\cite{scholz2016ratcheting}. Its motion closely resembles the active Brownian one~\cite{bechinger2016active, digregorio2018full, shaebani2020computational}, but with non-negligible inertia~\cite{Scholz2018inertial}. In contrast, the passive tracer (\autoref{Fig:Fig1}~A) has much thicker and symmetrical legs, providing no net self-propulsion by this principle, acting instead as an underdamped Brownian particle.
The radius $R$ of the passive tracer is chosen as twice the radius $r$ of the active bath particles, i.e.\ $R=2r$. In our study, we vary the packing fraction $\phi=r^2N_\text{a}/\mathcal R^2$ of the environment, where $N_\text{a}$ is the number of active particles and $\mathcal R$ is the radius of the circular plate, determining the system size.\par
Snapshots of the experiment are reported for $\phi=0,0.15,0.3$ in \autoref{Fig:Fig1}~D-F, corresponding $\sim\unit[80]{s}$ long
trajectories of the tracer are displayed in \autoref{Fig:Fig1}~G-I, and recordings of the system's full dynamics are shown in supplementary movies 1-3, respectively. A single isolated tracer driven by the shaking plate performs a random isotropic motion that mimics inertial Brownian motion with a diffusion constant $D_\text{t}$. In the presence of an active bath, and (to a point) as said bath's packing fraction $\phi$ grows, the passive tracer explores an ever larger area in the same time interval, indicating an enhanced diffusivity. Intuitively this can be explained by the increasing number of collisions with the active particles that transfer additional energy and momentum to the tracer. However, upon close inspection, the observed scenario differs significantly from a simple enhancement of $D_\text{t}$: The trajectories of the tracer display an increasing degree of persistence for short times (zooms in \autoref{Fig:Fig1}~G-I) and approach a diffusive motion only on longer timescales. The passive tracer thus performs a persistent motion typical of self-propelled objects and behaves as an active unit whose inertia cannot be neglected. This effect is purely induced by the non-equilibrium (active) bath with no counterpart in the equilibrium (passive) case.\par

\begin{figure*}
	\includegraphics[width=.85\textwidth]{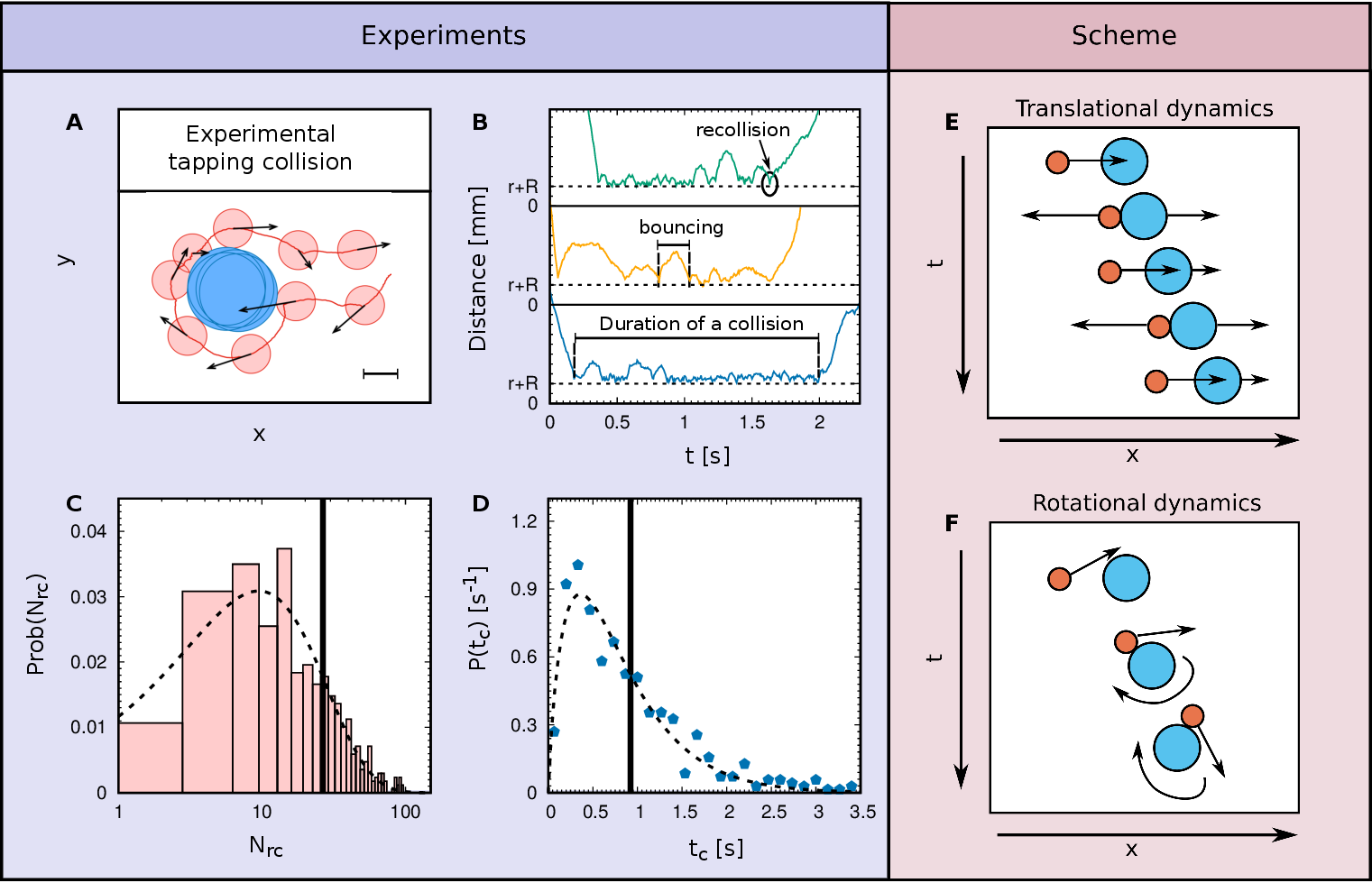}
	\caption{
		\label{Fig:Fig3}
		\textbf{Collision between active particles and passive tracer.}
		(\textbf{A}) Time evolution of a tapping collision between an active vibrobot (orange disk) and the passive tracer (light blue disk) in the $x,y$ plane.
		Black arrows denote the velocity of the active particle while the red line tracks its trajectory. The scale bars reads $\unit[15]{mm}$.
		(\textbf{B}) Relative distance between an active particle and the passive tracer as a function of time, $t$, for three different trajectories. The time is shifted to focus on the beginning of one collision event in the three cases. Recollisions, bouncing events, and the total duration of the collision are marked in each graph.
		(\textbf{C}) probability $ \text{Prob}(N_\text{rc})$ of observing a number $N_\text{rc}$ of recollisions during a collision.
		(\textbf{D}) probability $ P(t_\text{c})$ of observing a collision lasting a time $t_\text{c}$.
		Both in (\textbf{C}) and (\textbf{D}), dashed black lines are obtained by fitting the function $f(x) \propto e^{-x/a} x^{b}$ where $a$ and $b$ are fitting parameters.
		The vertical black lines in (\textbf{C}) and (\textbf{D}) mark the average $\langle N_\text{rc} \rangle$ and $\langle t_\text{c}\rangle$, respectively.
		(\textbf{E}-\textbf{F}) schemes of the collisions, showing the tapping scenario.
		(\textbf{E}) and (\textbf{F}) outline the persistent transfer of translational and rotational momentum from the active to the passive particle, respectively. 
		(\textbf{A}-\textbf{D}) are obtained from experimental data with $\phi=0.075$, while (\textbf{E}-\textbf{F}) are schematic illustrations. Therefore, in the latter case, arrows and disks are fictitious representation of particles' velocities and positions to highlight the effects of tapping collisions.
		Active and passive particles are represented by orange and blue disks, respectively.
	}
\end{figure*}

\subparagraph{Persistent tapping collisions.}
In a thermal background at equilibrium, particles of the environment collide, transfer momentum to the tracer, and depart again.  The random kicks provided by bath particles induce Brownian motion on the tracer this way. In an active environment, this scenario is violated, because particles are characterized by a degree of persistence that allows them to perform non-instantaneous interaction events. 
In the absence of inertia, active particles attach to walls and obstacles for a typical time~\cite{fily2014dynamics,vladescu2014filling} and, consequently, an overdamped active particle would persistently push the tracer before moving away. In contrast, here we show that the interplay between inertia and activity of the bath particles induces several recollisions during these non-instantaneous particle-particle interactions. We refer to this phenomenon as \emph{tapping collisions}. These also manifest as repeated bouncing events observed during the accumulation of inertial active particles near walls~\cite{Lam2015,leoni2020surfing}.\par
Our experiments provide a controlled setting to study temporally resolved tapping collisions and their consequences on the tracer dynamics.
\autoref{Fig:Fig3}~A shows one such typical event in the trajectory of an active vibrobot in the vicinity of the passive one, obtained from experimental data for $\phi=0.075$. In each individual impact, the active vibrobot exchanges momentum with the tracer and bounces back. Then it swiftly recovers due to its own activity and accelerates once more roughly towards the target, likely to collide with the tracer again. This picture is confirmed in \autoref{Fig:Fig3}~B, where we show measured distances between the passive tracer and three individual active particles as a function of time. These recollisions prolong the duration of particle-tracer interactions, resulting in what can be effectively interpreted as \emph{persistent collisions}. To quantify this phenomenon, \autoref{Fig:Fig3}~C displays the empirical probability $\text{Prob}(N_\text{rc})$ that an active particle experiences $N_\text{rc}$ impacts with the tracer during a single tapping collision. \autoref{Fig:Fig3}~D shows the empirical probability $P(t_\text{c})$ that a total persistent collision event of a character like shown in \autoref{Fig:Fig3}~B should last for as long as $t_\text{c}$. Recollisions are frequent and the total duration of tapping collisions is significantly larger than that of microscopic collisions typical of granular matter~\cite{Scholz2018}. $\text{Prob}(N_\text{rc})$ peaks at $N_\text{rc}\sim10$, has an average value $\langle N_\text{rc}\rangle\approx20$, and a long tail. A similar shape is observed in $P(t_\text{c})$ with an average recollision time of $\langle t_\text{c}\rangle\approx\unit[0.9]{s}$.\par
In the trajectory reported in \autoref{Fig:Fig3}~A (and similar ones), we also observe a sliding or self-alignment of the active particle along the passive tracer: After a bouncing event, the active vibrobot aligns its velocity to the tracer surface as if it was subject to an effective alignment interaction~\cite{Lam2015}.
This is consistent with our initial picture illustrated in \autoref{Fig:Fig0}, and can be explained by frictional contact forces between the particles.
Such forces would cause a torque on the active particle inducing its reorientation with respect to the surface of the passive tracer~\cite{brilliantov2007translations}.\par
Summarizing, in contrast to the typical quasi-instantaneous elastic (or non-elastic) collisions between passive particles, the phenomenon of tapping collisions shows two characteristic, non-trivial elements: Persistence and effective alignment.
Partial inelasticity of the collisions does not play a crucial role. As long as a sufficient fraction of total momentum is conserved, at most the number of bouncing events is reduced through dissipation. The tapping phenomenon has important consequences on the resulting tracer motion, as we elaborate in the following.\par

\subparagraph{Momentum transfer during a tapping collision.}
The right hand panels in \autoref{Fig:Fig3} schematically show the effect of persistent tapping collisions on the tracer dynamics: \autoref{Fig:Fig3}~E shows how multiple recollision events induce an effective activity on the translational dynamics of the tracer. The numerous bouncing events (recollisions) from an active particle moving at a constant speed to the passive tracer induce repeated tapping kicks from roughly similar directions. Each transfers translational momentum from the active to the passive particle. In this way, despite the presence of an internal friction force that dissipates part of the kinetic energy, the passive tracer nevertheless accumulates momentum and displays a net persistent motion. 
\autoref{Fig:Fig3}~F shows how the large size of the tracer allows active particles that come in for a strafing impact to transfer angular momentum during each collision as well. Each recollision then contributes to a similarly accumulating effective torque on the passive tracer, accelerating its rotational motion.\par 
These two effects crucially affect both the short and, quite surprisingly, the long time behavior of the passive tracer. In the following sections, we take a closer look at the statistical properties of the trajectories and illustrate the occurrence of persistent long time correlations. This has important consequences on the relation between the effective diffusion coefficient of the tracer and the stationary velocity distribution of the active bath, whence we derive a generalized active Einstein relation.\par

\subparagraph{Translational dynamics of the tracer.}
In contrast to the bath's active vibrobots, the ``activated'' tracer does not move at constant speed nor aims to move in a definite direction with respect to its orientation.
Therefore, a common model such as active Brownian particles~\cite{bechinger2016active, shaebani2020computational} is not suited to describe its dynamics.
Here, we demonstrate by experiments and theory - through an explicit coarse-graining of the microscopic dynamics - that the translational motion of the tracer can be accurately modelled by the inertial active Ornstein-Uhlenbeck particle (AOUP) dynamics heuristically introduced in Refs.~\cite{caprini2021inertial, nguyen2021active}.\par
\begin{figure*}
	\includegraphics[width=\textwidth]{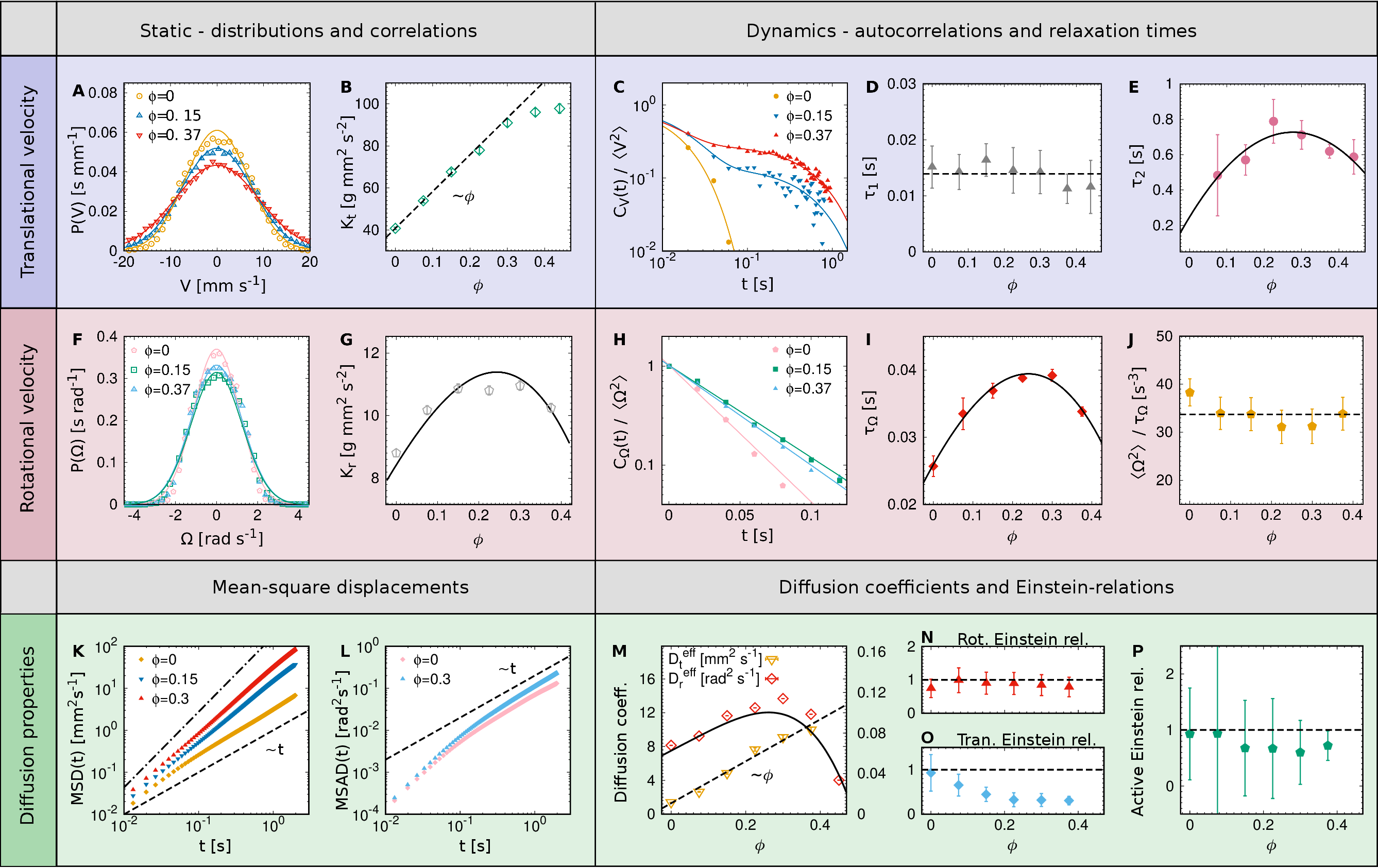}
	\caption{
		\label{Fig:tracerobservables}
		\textbf{Statics and dynamics of the tracer.}
		(\textbf{A}) Probability distribution $P(V)$ of a Cartesian component of the velocity for several values of the packing fraction $\phi$.
		(\textbf{B}) Kinetic energy $\mathcal{K}_\text{t}=M\langle\mathbf{V}^2\rangle/2$ as a function of $\phi$.
		(\textbf{C}) Autocorrelation of the velocity $C_V(t)$ for several values of $\phi$.
		(\textbf{D}-\textbf{E}) Typical times of $C_V(t)$ calculated from the profile $\alpha e^{-t/\tau_1}+\beta e^{-t/\tau_2}$ as a function of $\phi$.
		(\textbf{F}) Probability distribution of the angular velocity $P(\Omega)$ for several values of $\phi$.
		(\textbf{G}) Rotational kinetic energy $\mathcal{K}_\text{r}=J\langle\Omega^2\rangle/2$ as a function of $\phi$.
		(\textbf{H}) Angular velocity autocorrelation $C_\Omega(t)$ for several values of $\phi$.
		(\textbf{I}) Typical time of $C_\Omega(t)$, extracted from the profile $\sim e^{-t\tau_\Omega}$ as a function of $\phi$.
		(\textbf{J}) Ratio $\langle\Omega^2\rangle\tau_{\Omega}$ as a function of $\phi$.
		(\textbf{K}-\textbf{L}) Mean-square displacement $\text{MSD}(t)=\langle\left(\mathbf{X}(t)-\mathbf{X}(0)\right)^2\rangle$, and mean-square angular displacement $\text{MSAD}(t)=\langle\left(\Phi(t)-\Phi(0)\right)^2\rangle$, for several values of $\phi$.
		(\textbf{M}) Diffusion coefficients, $D_\text{t}^\text{eff}$ (left y-scale) and $D_\text{r}^\text{eff}$ (right y-scale), as a function of $\phi$. The units of $D_\text{t}^\text{eff}$ and $D_\text{r}^\text{eff}$ are $\unit[1.18]{mm^2s^{-1}}$ and $\unit[1.18]{rad^2s^{-1}}$, respectively.
		(\textbf{N}) Einstein relations for the rotational degree of freedom, $\langle\Omega^2\rangle\tau_\Omega/D_\text{r}^\text{eff}$.
		(\textbf{O}) Conventional Einstein relation for the translational degree of freedom, $\langle\mathbf{V}^2\rangle M/(\Gamma_\text{r}D_\text{t}^\text{eff})$.
		(\textbf{P}) Active Einstein relation for $D_\text{t}^\text{eff}$, given by \autoref{eq:Generalized_Einstein_relation}.
		In all the panels, points reflect experimental data, solid colored lines are exponential fitting functions justified by our coarse-grained models, black dashed lines denote linear or constant scaling, black dashed-dotted lines evidence ballistic behavior, and solid black lines are guides for the eyes.
		In (\textbf{B}), (\textbf{D}), (\textbf{E}), (\textbf{G}), (\textbf{I}), (\textbf{J}), (\textbf{M}-\textbf{P}), error bars represent standard deviation.
	}
\end{figure*}

In \autoref{Fig:tracerobservables}~A, the steady-state distribution $P(V)$ of a Cartesian component of the velocity $V=V_x$ is plotted for several values of the packing fraction $\phi$. The shape of $P(V)$ is described by a Gaussian profile, both with and without a bath. By contrast, active vibrobots are characterized by a double peak distribution, caused by their typical speed induced by the active force~\cite{Scholz2018inertial}. This difference signifies that our tracer cannot be described by an inertial active Brownian model while the inertial AOUP is a more suitable dynamics to reproduce its behavior (see also the Methods). 
Here, collisions broaden $P(V)$, inducing a monotonic increase of the translational kinetic energy of the tracer, $\mathcal{K}_\text{t}=M\langle\mathbf{V}^2\rangle/2$, where $M$ is the mass of the passive tracer (see \autoref{Fig:tracerobservables}~B). Intuitively, a larger packing fraction $\phi$ means more collisions per unit of time which transfers more energy from the bath to the tracer, thus enhancing its translational motion. For small $\phi$, the behavior of $\mathcal{K}_\text{t}(\phi)$ can be understood through a scaling argument
\begin{equation}\label{eq:scaling}
	\mathcal{K}_\text{t}-D_\text{t}\Gamma_\text{t}\sim\phi\,,
\end{equation}
where $\Gamma_\text{t}$ and $D_\text{t}$ are the friction and translational diffusion coefficients of the tracer, and thus $D_\text{t}\Gamma_\text{t}$ corresponds to its kinetic temperature in the absence of active particles ($\phi=0$). We refer to the average kinetic energy as \emph{temperature} here in analogy to its thermodynamic counterpart. However, since the origin of fluctuations in our system is athermal, it depends only on the kinetic degrees of freedom of the tracer particle and their excitation from the vibrating plate, independently of the actual ambient conditions.
Details on this scaling relation are reported in the methods section. A deviation from the linear scaling is observed at packing fractions $\phi\gtrsim0.3$. At these high densities, many active particles collide simultaneously with the tracer from various directions (see Supplementary Movie 3), so that their impacts balance out and only induce additional dissipation into the tracer's internal degrees of freedom.\par
To characterize the tracer's translational motion we evaluate the velocity autocorrelation $C_V(t)=\langle\mathbf{V}(t)\cdot\mathbf{V}(0)\rangle$. As seen in \autoref{Fig:tracerobservables}~C, $C_V(t)$ shows a single exponential decay regime for $\phi=0$ as in equilibrium, but two distinct decay regimes for $\phi>0$. The short time regime does not depend on $\phi$ (\autoref{Fig:tracerobservables}~D) and is determined by the passive tracer's properties ($\phi=0$). With the translational friction coefficient $\Gamma_\text{t}$ the characteristic decay time is $\tau_1\approx M/\Gamma_\text{t}$. The long time regime is induced by the active bath and depends non-monotonically on the packing fraction as evident when plotting its typical time $\tau_2$ against $\phi$, as we did in \autoref{Fig:tracerobservables}~E. 
As $\tau_2$ is determined by the inverse of the active vibrobots' rotational diffusion coefficient, it is an order of magnitude larger than the inertial time of the tracer, i.e.\ $\tau_2\sim 1/D^\text{a}_\text{r}\approx10\tau_1$.
The full expression for $\tau_2$ accounting for the effect of the rotational inertia is reported in the methods (see Eq.~ \eqref{eq:tau_effective_rotinertia}).
Intuitively, the increase in the number of collisions is responsible for the enhancement of the persistence time preventing the tracer to rapidly dissipate energy gained from any given collision. However, at larger $\phi\gtrsim0.3$ this effect is counteracted by simultaneous collisions with different bath particles, which randomize the direction of the tracer's net motion and lower $\tau_2$.\par
By approximating the volume exclusion interaction between the passive tracer and active vibrobots as %by
harmonic springs, we can derive the tracer's effective translational dynamics, coarse-graining the bath's degrees of freedom by a non-Markovian noise term. As shown in the methods section, this results in an underdamped coarse-grained dynamics
\begin{equation}\label{eq:effectivedynamics_translational}	M\dot{\mathbf{V}}+\Gamma_\text{t}\mathbf{V}+\Gamma_\text{t}\sqrt{2D_\text{t}}\boldsymbol\xi=\Gamma_\text{t}V_0\boldsymbol\eta
\end{equation}
known in the literature as inertial active Ornstein-Uhlenbeck particle (AOUP). In this model, $\boldsymbol\xi$ is a white noise term with zero average and unit variance. The friction coefficient $\Gamma_\text{t}$, and the diffusion coefficient $D_\text{t}$ are measured in the absence of the active bath and, in the experiment, are due to friction with the substrate and the shaker's oscillation amplitude. 
Here we model the friction as a Stokes force rather than a constant one. Indeed, for a similar setup~\cite{Scholz2018}, the velocity of a kicked vibrobot has consistently revealed an exponential temporal decay rather than a simple linear decrease.
The persistent tapping collisions introduce an additional, non-Markovian noise term, $\Gamma_\text{t}V_0\boldsymbol\eta$, which provides to the tracer a typical speed $V_0(\phi)$, depending on the packing fraction and the active vibrobots' inherent self-propulsion. The stochastic term $\boldsymbol\eta$ is an Ornstein-Uhlenbeck process with characteristic time $\tau_2$ and autocorrelation
\begin{equation}
	\langle\boldsymbol\eta(t)\boldsymbol\eta(0)\rangle\sim e^{-t/\tau_2}\,.
\end{equation}
Our coarse-grained model \eqref{eq:effectivedynamics_translational} for the translational dynamics of the tracer can be solved analytically and reproduces the experiments' static and dynamic observables to within statistical error. Indeed, it gives rise to a Gaussian velocity distribution and a velocity autocorrelation $C_V(t)$ characterized by the sum of two exponentials
\begin{equation}\label{eq:CV}
	C_V(t)\sim e^{-t/\tau_1}+Ae^{-t/\tau_2}\,,
\end{equation}
where $A$ is a coefficient depending on the parameters of the system (see details in the methods section).\par

Since our theoretical findings are consistent with the experimental results, the effective translational motion of the tracer in our experiments constitutes the first experimental realization of the inertial AOUP~\cite{caprini2021inertial, nguyen2021active}. Here, the activity originates directly from a non-Markovian memory kernel $\boldsymbol\eta$ in the equation of motion, which arises from the tapping collisions with the particles of the active bath. In contrast to active Brownian motion, here there is no coupling between orientational and translational degrees of freedom.

\subparagraph{Rotational dynamics of the tracer.}
To quantify the effect of the active bath on the tracer's rotational dynamics, we analyze the distribution $P(\Omega)$ of the angular velocity $\Omega$. As shown in \autoref{Fig:tracerobservables}~F, $P(\Omega)$ has an almost-Gaussian shape for several values of the packing fraction $\phi$, with only minor deviations, which are expected for driven dissipative granular systems~\cite{Scholz2017, yu2020velocity, eshuis2010experimental}. Here, we neglect these effects, and assume that the motion is sufficiently well-described by a Gaussian profile, and fully characterized by the tracer's rotational kinetic energy $\mathcal{K}_\text{r}=J \langle\Omega^2\rangle/2$ with moment of inertia $J$.
$\mathcal{K}_\text{r}(\phi)$ displays a non-monotonic profile as a function of $\phi$ (see \autoref{Fig:tracerobservables}~G). In the absence of active particles ($\phi=0$), we find that $\mathcal{K}_\text{r}(0)\sim D_\text{r}\Gamma_\text{r}/J$, where $\Gamma_\text{r}$ and $D_\text{r}$ are the isolated tracer's rotational friction and diffusion coefficients, respectively. The number of tapping collisions grows when $\phi$ increases. Much like in the case of translation, angular momentum, too, is transferred to the tracer, as described in \autoref{Fig:Fig3}~F, and should intuitively be proportional to the number of collisions in the dilute regime ($0\leq\phi\lesssim0.3$). In contrast, for $\phi\gtrsim0.3$, simultaneous collisions occur and balance each other by providing clockwise and counterclockwise torques, leaving only friction to reduce the tracer's net rotational kinetic energy.\par
The autocorrelation $C_\Omega(t)=\langle\Omega(t)\Omega(0)\rangle$ of angular velocity, shown in \autoref{Fig:tracerobservables}~H for different values of $\phi$, displays an exponential-like shape with typical time $\tau_\Omega$, which can be interpreted as the ratio between the moment of inertia $J$ and an effective rotational friction coefficient $\Gamma_\text{r}^{\text{eff}}$. As seen in \autoref{Fig:tracerobservables}~I, $\tau_\Omega$ depends non-monotonically on the packing fraction $\phi$. For $\phi=0$, the case of passive Brownian motion is recovered and $\tau_\Omega(0)\sim J/\Gamma_\text{r}$. 
In the dilute regime ($\phi\lesssim0.3$), the autocorrelation time $\tau_\Omega$ increases as a function of $\phi$, because rotational velocity is accumulated during a tapping collision. 
In particular, this effect is induced by the sliding motion of the vibrobots around the passive tracer. 
The active bath enhances the tracer particle's effective rotational inertia up to twice its passive value. Consequently, the tracer evolves as if subject to an effective rotational friction coefficient $\Gamma_\text{r}^\text{eff}<\Gamma_\text{r}$. Finally, for $\phi\gtrsim0.3$, simultaneous collisions hinder this effect reducing $\tau_\Omega$ and, thus, increasing $\Gamma_\text{r}^\text{eff}$.\par
Since rotational dynamics are only determined by a single relevant timescale, with almost Gaussian properties and exponential autocorrelation, an appropriate theoretical description consists of an underdamped Langevin equation of the form
\begin{equation}
	J\dot\Omega=-\Omega\Gamma_\text{r}^\text{eff}+\Gamma_\text{r}^\text{eff}\sqrt{2D^\text{eff}_\text{r}}\xi_\text{r}\,,
	\label{eq:effectivedynamics_rotational}
\end{equation}
where $\xi_\text{r}$ is a white noise term with zero average and unit variance and $D^\text{eff}_\text{r}$ represents an effective rotational diffusion coefficient. To provide additional information on the non-monotonic dependence on the packing fraction, \autoref{Fig:tracerobservables}~J shows the experimental ratio between $\mathcal{K}_\text{r}$ and $\tau_\Omega$ as a function of $\phi$. We find that $\mathcal{K}_\text{r}(\phi)/\tau_\Omega(\phi)\approx\mathcal{K}_\text{r}(0)/\tau_\Omega(0)$ is constant, i.e.\ the rotational kinetic energy and rotational time have the same dependence on $\phi$. Since $\mathcal{K}_\text{r}/\tau_{\Omega}=D_\text{r}^\text{eff}(\Gamma^{\text{eff}}_\text{r})^2/J$, the effective rotational diffusion coefficient scales as $D_\text{r}^\text{eff}\propto1/(\Gamma^{\text{eff}}_\text{r})^2\propto\tau_\Omega^2$.\vskip\baselineskip\par
To summarize, the effective inertial AOUP motion~\eqref{eq:effectivedynamics_translational} for the translational dynamics is accompanied by an intrinsic rotational motion~\eqref{eq:effectivedynamics_rotational} enhanced by tapping collisions. This effect is due to the tangential motion of active particles along the surface of the tracer that leads to additional rotation in its dynamics.

\begin{figure*}
	\includegraphics[width=0.9\textwidth]{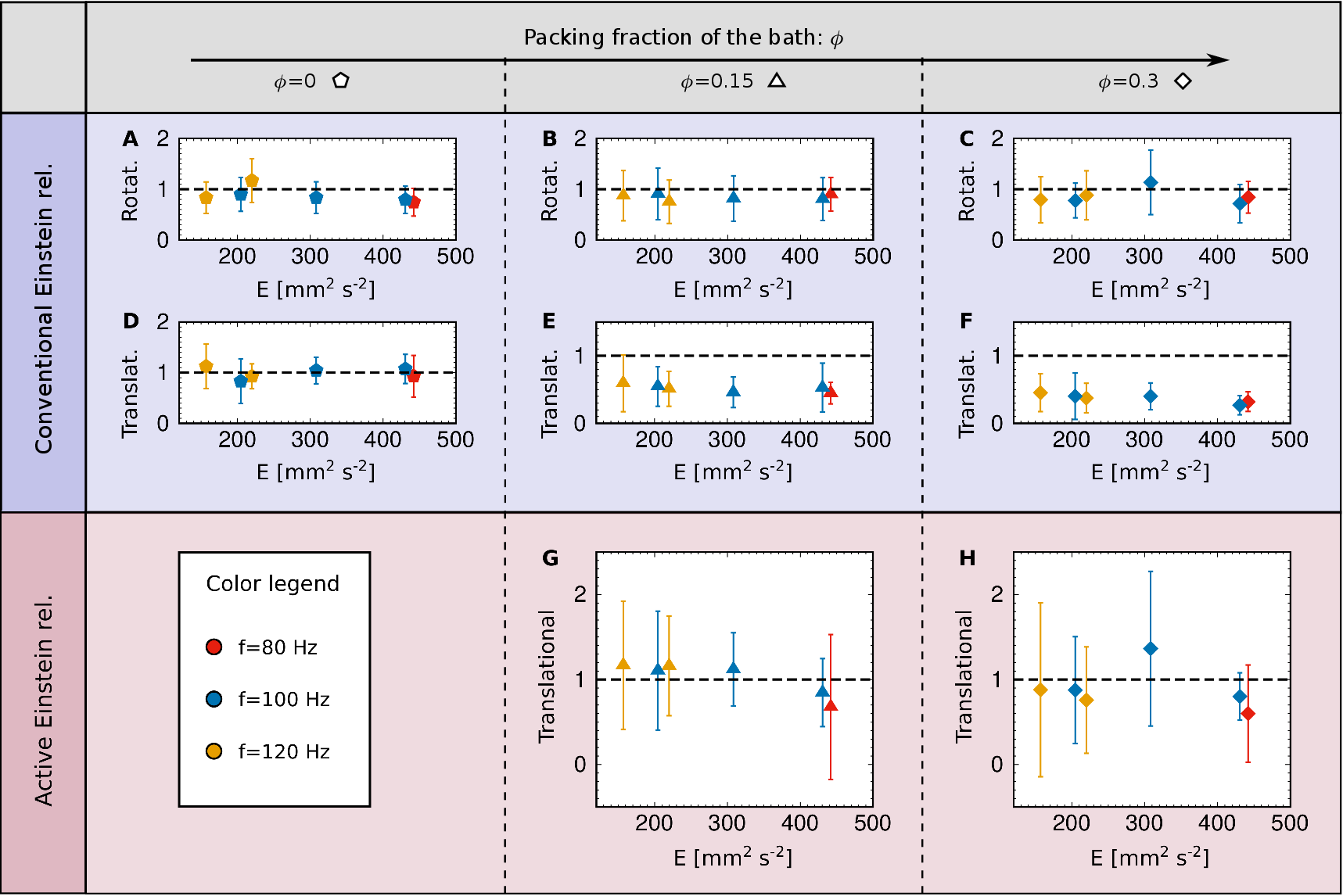}
	\caption{
		\label{Fig:Fig6}
		\textbf{Conventional and generalized Einstein relations for different shaker settings}. Einstein relation are shown as a function of the energy injected by the shaker $E=(2\pi f)^2 A^2/2$, where $\unit[f]{[Hz]}$ and $\unit[A]{[mm]}$ are frequency and amplitude of the oscillation of the shaker, respectively. As indicated by the horizontal axis, we consider three different packing fractions of the active bath: $\phi=0$ (only the passive tracer), $\phi=0.15$, and $\phi=0.3$.
		(\textbf{A}-\textbf{C}) Einstein relation for the rotational degree of freedom, $\langle\Omega^2\rangle\tau_\Omega/D_\text{r}^\text{eff}$.
		(\textbf{D}-\textbf{F}) Conventional Einstein relation for the translational degree of freedom, $\langle\mathbf{V}^2\rangle M/(\Gamma_\text{r}D_\text{t}^\text{eff})$.
		(\textbf{G}-\textbf{H}) Generalized active Einstein relation for $D_\text{t}^\text{eff}$, given by \autoref{eq:Generalized_Einstein_relation}.	
		In all the panels, points reflect experimental data and black dashed lines are eye guides marking one, i.e.\ the validity of the Einstein relations.
		Error bars represent one standard deviation. 
	}
\end{figure*}

\subparagraph{Active Einstein relations.}
In equilibrium, a particle's diffusion coefficient is determined by its thermal energy and mobility via the Einstein-Sutherland-Smoluchowski relation~\cite{marconi2008fluctuation, baldovin2022many}. This result implies a deep connection between fluctuations which induce diffusion, and dissipation due to the friction coefficient, i.e.\ the inverse of mobility. Out of equilibrium the Einstein relation is usually violated, because the diffusion is not only determined by thermal fluctuations but also by additional non-equilibrium mechanisms. In our system we expect that such a violation is caused by the extra source of energy transferred from the environment to the tracer through persistent tapping collisions.\par
While the Einstein relation is also violated in our experiments, the shape of the autocorrelation functions and our effective description allow us to reconcile the relation with the notion of emergent activity. To this end, we derive and experimentally verify generalized active Einstein relations for rotational and translational tracer dynamics: Both equilibrium and non-equilibrium fluctuations of the system, stored as translational and rotational kinetic energy, determine diffusive properties of the tracer through friction coefficients (related to equilibrium dissipation) and characteristic persistence times (quantifying the departure from equilibrium~\cite{fodor2016far}).\par
To construct our active Einstein relations, we first calculate the mean squared displacement $\text{MSD}(t)=\langle\left(\mathbf X(t)-\mathbf X(0)\right)^2\rangle$ and the mean squared angular displacement, $\text{MSAD}(t)=\langle\left(\Phi(t)-\Phi(0)\right)^2\rangle$ from the tracer positions $\mathbf X(t)$ and orientations $\Phi(t)$, as shown in \autoref{Fig:tracerobservables}~K and \autoref{Fig:tracerobservables}~L, respectively. From the linear behavior in the long time limit, we measure and display in \autoref{Fig:tracerobservables}~m the effective translational and rotational diffusion coefficients $D_\text{t}^\text{eff}$ and $D_\text{r}^\text{eff}$ under the influence of the active bath. As the packing fraction $\phi$ grows, $D_\text{t}^\text{eff}$ shows a monotonic increase. Although collisions always enhance the rotational diffusivity with respect to $\phi=0$, $D_\text{r}^\text{eff}$ qualitatively displays a similar non-monotonicity as $\tau_\Omega$ (see \autoref{Fig:tracerobservables}~I). We can thereby identify an optimal $\phi$ that maximizes the angular diffusion properties.\par
From the similar behavior of $D_\text{r}^\text{eff}$ and $\tau_\Omega$ we suggest the Einstein relation
\begin{equation}\label{eq:Rot_Einstein_rel}
	D_\text{r}^\text{eff}=\langle\Omega^2\rangle\tau_\Omega
\end{equation}
for the rotational dynamics, which is experimentally validated in \autoref{Fig:tracerobservables}~N for the values of $\phi$ analyzed in this work. The enhanced angular fluctuations induced by collisions cause an effective rotational diffusivity that is consistent with the scenario of an effective angular temperature depending on $\phi$. However, since our system is entirely driven athermally, this effective temperature is not related to the thermal properties of the bath. 
On the contrary, the analogous conventional Einstein relation for the translational diffusivity is strongly violated as $\phi$ is increased. Indeed, in \autoref{Fig:tracerobservables}~O, we observe $D_\text{t}^\text{eff}\geq\langle\mathbf{V}^2\rangle M/\Gamma_\text{t}$. This violation is due to the additional correlated injection of energy to the tracer dynamics by tapping collisions. 
Additionally, it is a dynamical signature of non-equilibrium effects, such as the second decay regime in the time shape of $C_V(t)$. This also reflects the breakdown of the effective temperature scenario as a consequence of the non-equilibrium properties of the active bath.\par
Using the effective dynamics \eqref{eq:effectivedynamics_translational}, we derive the generalized active Einstein relation
\begin{equation}\label{eq:Generalized_Einstein_relation}
	D_\text{t}^\text{eff}=-D_\text{t}\tau_2\frac{\Gamma_\text{t}}{M}+\langle\mathbf{V}^2\rangle\left(\tau_2+\frac M{\Gamma_\text{t}}\right)
\end{equation}
expressing the kinetic fluctuations of the tracer as a function of the translational diffusion coefficient (see methods section for further details). This prediction is verified by experimental data in \autoref{Fig:tracerobservables}~P, revealing a good agreement within the statistical error. In contrast to the conventional Einstein relation, \autoref{eq:Generalized_Einstein_relation} involves not only equilibrium dissipation through the inertial time $\tau_1=M/\Gamma_\text{t}$, but also the collision-induced active force, through the persistence time $\tau_2\approx\frac1{D_\text{r}^\text{a}}\left(1+\frac{J_\text{a}}{\gamma_\text{r}}D^a_\text{r}\right)$. In other words, it accounts for both typical times governing the decay of the velocity autocorrelation $C_V(t)$. Our generalized Einstein formula \eqref{eq:Generalized_Einstein_relation} contains an equilibrium term $\langle\mathbf{V}^2\rangle M/\Gamma_\text{t}$ and two non-equilibrium terms $\langle\mathbf{V}^2\rangle\tau_2-D_\text{t}\tau_2M/\Gamma_\text{t}>0$. Indeed, they vanish in the equilibrium limit $\tau_2\to0$, so that the conventional Einstein relation $D_\text{t}=\langle\mathbf{V}^2\rangle M/\Gamma_\text{t}$ is recovered. The relevance for applications is that $\tau_2$ can be approximated by the persistence time of the active particle $\tau_2\approx\frac1{D_\text{r}^\text{a}}\left(1+\frac{J_\text{a}}{\gamma_\text{r}}D^a_\text{r}\right)$, where the correction is due to the rotational dynamics of the vibrobots.
which allows an estimate of the effective diffusion from the bulk behavior at low densities.\par

\begin{figure*}
	\includegraphics[width=\textwidth]{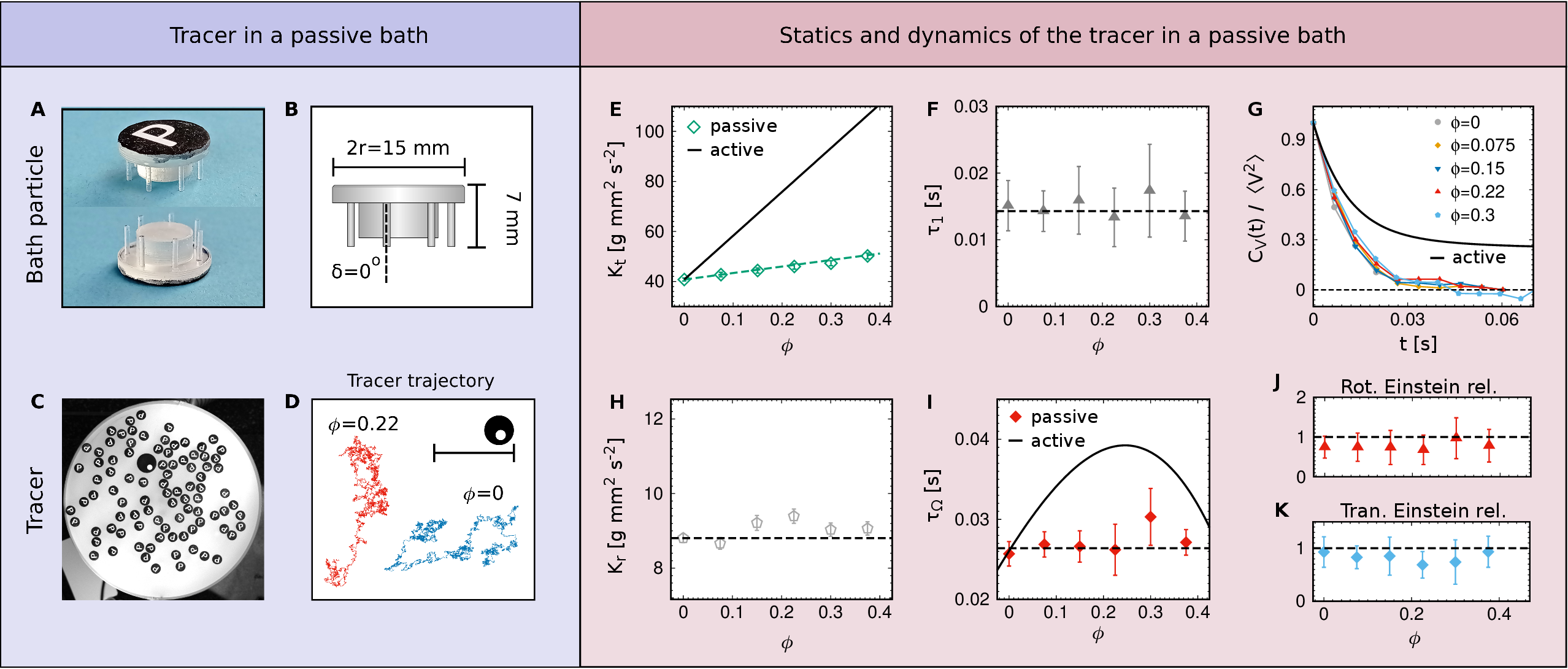}
	\caption{
		\label{Fig:appendix}
		\textbf{Statics and dynamics of the tracer in a passive bath.}
		(\textbf{A}-\textbf{B}) Photo and diagram of a passive bath particle (appropriately labeled with a capital \emph P, for \emph{Passive}).
		(\textbf{C}) Photo of the experiment for a tracer in a passive bath at packing fraction $\phi=0.22$.
		(\textbf{D}) Trajectories of the tracer for $\phi=0$ (blue) and $\phi=0.22$ (red). The scale bar reads $\unit[50]{mm}$.
		(\textbf{E}) Kinetic energy $\mathcal{K}_\text{t}=M\langle\mathbf{V}^2\rangle/2$ as a function of $\phi$. We also report as a black solid line the linear scaling of $\mathcal{K}_\text{t}$ for a tracer in an active bath, see \autoref{Fig:tracerobservables}~B.
		(\textbf{F}) Typical time $\tau_1$ of the autocorrelation of the velocity $C_V(t)$ calculated from the profile $\alpha e^{-t/\tau_1}$ as a function of $\phi$.
		(\textbf{G}) $C_V(t)$ for several values of $\phi$. Here, the solid black line marks the double exponential decay observed in \autoref{Fig:tracerobservables}~c for $\phi=0.15$.
		(\textbf{H}) Rotational kinetic energy $\mathcal{K}_\text{r}=J\langle\Omega^2\rangle/2$ as a function of $\phi$.
		(\textbf{I}) Typical time of angular velocity autocorrelation $C_\Omega(t)$, extracted from the profile $\sim e^{-t\tau_\Omega}$ as a function of $\phi$. 
		The solid black line marks the value of $\tau_\Omega$ as a function of $\phi$ in the presence of the active bath, see \autoref{Fig:tracerobservables}~{I}.
		(\textbf{J}) Einstein relation for the rotational degree of freedom, $\langle\Omega^2\rangle\tau_\Omega/D_\text{r}^\text{eff}$.
		(\textbf{K}) Conventional Einstein relation for the translational degree of freedom, $\langle\mathbf{V}^2\rangle M/(\Gamma_\text{r}D_\text{t}^\text{eff})$.
		In all the panels, points reflect experimental data and black dashed lines denote linear or constant scaling. 
		In (\textbf{E}-\textbf{F}), (\textbf{H}-\textbf{K}) error bars represent one standard deviation.
	}
\end{figure*}

\subparagraph{Validity of Einstein relation.}
To extend the generality of our results, we perform experiments by changing the frequency $f$ and the amplitude $A$ of the shaker, that control the energy $E=(2\pi f)^2 A^2/2$ (normalized by the total mass of the load) injected in the system.
The behaviour shown in \autoref{Fig:tracerobservables} is preserved qualitatively in all explored configurations: Active particles perform tapping collisions against the tracer, which in turn behaves as an active particle characterized by a persistent trajectory and described by the inertial active Ornstein-Uhlenbeck model. Indeed, the autocorrelation of translational and angular velocity maintains the same exponential structure, with two and one temporal decay regimes, respectively.\par
Our theoretical results and, in particular, the violation of the conventional Einstein relation, as well as the validity of the generalized active Einstein relation, are shown in \autoref{Fig:Fig6} as a function $E$ for three different values of the packing fraction of the active bath.
At $\phi=0$, in the absence of the active bath, the tracer behaves like a passive particle for every $E$ and thus the conventional Einstein relations for rotational (\autoref{Fig:Fig6}~A) and translational dynamics (\autoref{Fig:Fig6}~D) hold.
In contrast, for $\phi=0.15$ and $\phi=0.3$, the tracer behaves like an effective active Ornstein-Uhlenbeck particle with inertia.
As a result, the Einstein relation for the rotational dynamics holds with the corresponding effective rotational diffusion coefficient $D_\text{r}^\text{eff}$ (\autoref{Fig:Fig6}~B-C) while the conventional Einstein relation for the translational dynamics is strongly violated (\autoref{Fig:Fig6}~E-F), i.e.\ $D_\text{t}^\text{eff}\geq\left\langle\mathbf{V}^2\right\rangle M/\Gamma_\text{t}$.
The generalized active Einstein relation is verified for all combinations of ($f$, $A$) and $\phi>0$ extending the applicability of our theoretical results (\autoref{Fig:Fig6}~G-H).

\subparagraph{Difference between active and passive baths.}
To investigate the unique effects of the active bath on the tracer dynamics, we compare our results with those obtained in the presence of a passive bath at different packing fractions $\phi$. A passive bath is realized by replacing active vibrobots with passive particles, whose legs are perpendicular to the plane of motion (\autoref{Fig:appendix}~A-B). The particle's shape is fully symmetric and there is, by design, no significant preferential direction in the motion. These symmetric vibrobots behave as passive Brownian particles subject to inertia, without relevant persistence in the particle trajectory. 
As a consequence, bath particles generally depart from the tracer after a single impact and do not return for a tapping collision.
A tracer in a passive bath (\autoref{Fig:appendix}~C) displays a Brownian trajectory with negligible persistence that resembles that observed at $\phi=0$ (\autoref{Fig:appendix}~D). \par
The tracer's effective static and dynamic behaviours in the presence of a passive bath are investigated in \autoref{Fig:appendix}~E-K. In \autoref{Fig:appendix}~E and \autoref{Fig:appendix}~H, the tracer's translational kinetic energy $\mathcal{K}_\text{t}$ and the rotational kinetic energy $\mathcal{K}_\text{r}$ are studied each as a function of the bath's packing fraction $\phi$. In both cases, we observe a much weaker dependence on $\phi$ than in corresponding experiments realized with an active bath. The momentum transferred during a collision between the tracer and a passive bath particle is less energetic because the typical speed of passive particles is smaller than the speed of their active counterparts. The speed of passive particles is indeed determined solely by random fluctuations.\par
In the presence of the passive bath, the translational velocity autocorrelation displays only a single decay regime with an exponential shape (\autoref{Fig:appendix}~G). This result qualitatively contrasts with that obtained in the presence of the active bath, where two decay regimes are observed. 
The single decay time $\tau_1$ measured from the velocity autocorrelation of the tracer in the passive bath does not depend on $\phi$ (\autoref{Fig:appendix}~F), coincides with $M/\Gamma$, and, thus, with the first decay regime measured in the presence of the active bath (\autoref{Fig:tracerobservables}~D). The single exponential behavior in the angular velocity autocorrelation is replicated also in the presence of a passive bath (not shown). In contrast to the experiments with the active bath, here the typical time $\tau_\Omega$ does not depend on $\phi$ (\autoref{Fig:appendix}~I) and, thereby, $\tau_\Omega\approx J/\Gamma_\text{r}$.\par
We conclude that the active bath is able to induce unique properties on the dynamics of the tracer with no equilibrium counterparts. Unlike the active bath, the passive bath is only able to change the static properties of the tracer, such as kinetic energies and diffusion coefficients, but cannot affect its dynamical properties, e.g.\ the temporal decay of autocorrelations. Consequently, the equilibrium form of the Einstein relation is recovered for the rotational dynamics, i.e.\ \autoref{eq:Rot_Einstein_rel} (\autoref{Fig:appendix}~J), and for the translational dynamics, i.e.\ the special case of \autoref{eq:Generalized_Einstein_relation} with $\tau_2=0$ (\autoref{Fig:appendix}~K).

\section{Discussion}
In conclusion, we have provided direct experimental observation of the full translational and rotational dynamics of a passive tracer moving in a non-equilibrium environment consisting of circle-shaped quasi-rigid active particles. We demonstrated, that an active bath induces persistent correlations and memory effects via collisions, that fundamentally differ from those characterizing standard passive liquids. The tracer's behavior cannot be fully described by an equilibrium model due to the emergence of additional relaxation times. The transfer of energy and momentum due to collisions is based on bounce-back effects and effective alignment: A new interaction mechanism, termed \emph{tapping collisions} is introduced to account for it, and allows us to derive a generalized active Einstein relation, linking fluctuations, dissipation, and activity.\par
Previous experimental studies, based on passive objects in a bath of bacteria or eukaryotic swimmers, mainly focus on the tracer position, reporting the enhancement of diffusivity~\cite{PhysRevLett.84.3017, PhysRevLett.106.018101, PhysRevLett.103.198103, kurtuldu2011enhancement}, the spontaneuous rotations of asymmetric micro-gears~\cite{di2010bacterial, sokolov2010swimming}, or the generalization of the energy equipartition theorem~\cite{maggi2014generalized}. With these microscopic experimental setups, only the violation of the Einstein relation was experimentally investigated in a bath of bacteria~\cite{PhysRevLett.99.148302, maggi2017memory}.
Indeed, a constructive derivation of an explicit relation in terms of direct experimental observables could not be easily obtained for such systems, because the short time dynamics cannot be resolved down to the particle level in microscopic systems, as positions or velocities of individual bacteria or colloids can be difficult to measure.
Consequently, dynamical information, such as translational and rotational velocity autocorrelations could usually not be fully investigated, and therefore the transfer of energy from the non-equilibrium bath to the tracer remained poorly understood. In contrast, our experimental setup, based on macroscopic vibrobots, allows us to investigate the problem in greater detail. We have provided a particle-resolved description for the collisions with direct measurements of the velocities of tracer and active particles. 
%This approach was necessary to derive the generalized active Einstein relation and could have profound consequences for biomedical applications, chemical applications, and swarms robotics.\par
This approach was necessary to derive the generalized active Einstein relation and could have consequences for two-dimensional swarm robotics applications. \par
A passive tracer in a bath of active vibrobots behaves as an underdamped active particle, whose inertia cannot be neglected. Tapping collisions with active vibrobots are responsible for the persistent transfer of translational and rotational momentum to the tracer, inducing persistent motion with increasing velocity. 
The active bath affects both static and dynamic properties of the tracer, in contrast to the passive bath, which leaves the tracer's dynamics unchanged.
The effective dynamical properties of the tracer, such as persistence time and effective rotational friction coefficient, are maximized by optimal values of the packing fraction of the bath, rather than stay monotonic with it.\par
Our study also provides an experimental realization of a model heuristically introduced in the framework of macroscopic active matter with inertia, termed inertial active Ornstein-Uhlenbeck particle (AOUP)~\cite{caprini2021inertial, nguyen2021active}.
Due to its simplicity and resulting theoretical possibilities, the inertial AOUP model plays an important role in the comprehension of inertial active systems providing theoretical advances to the physics of macroscopic active matter~\cite{lowen2020inertial}.
Here, the inertial AOUP model is derived by explicit coarse-graining of bath particle dynamics and its properties are fundamental to describe experimental data of athermal macroscopic systems dominated by inertia.

\section{Methods}
\subsection{Experimental details}
\subparagraph{Particle fabrication.}
Active vibrobots consist of a cylindrical core with a diameter of \unit[9]{mm} and a height of \unit[4]{mm} and a cylindrical cap ontop thereof, with a diameter of \unit[15]{mm} and a height of \unit[2]{mm}. This cap is supported by seven tilted cylindrical legs attached to it, each with a diameter of \unit[0.8]{mm}, and arranged in a regular heptagon around the core, elevating the cap's bottom face to \unit[5]{mm} above the substrate, and the core's but \unit[1]{mm}. The particle's total height is therefore \unit[7]{mm}. The seven legs are tilted to make an angle or $\delta=4^{\circ}$ with the vertical. The mass of the active vibrobots is $m=\unit[0.83]{g}$ while the moment of inertia is approximately $J_\text{a}=\unit[17.9\cdot10^{-8}]{g\,mm^2}$.\par
The passive vibrobot consists of an open cylindrical cap with an outer diameter of \unit[30]{mm}, an inner diameter of \unit[17.2]{mm}, and height of \unit[2]{mm}. Seven half-ellipsoidal legs with minor radius $\unit[1.5]{mm}$ and major radius $\unit[5]{mm}$ are attached to the cap so that the passive tracer has the same height as the active vibrobots. In contrast to the active particles' legs, the passive tracer's are by design vertical. They are much thicker, and have a rounded tip, so as to minimize the impact of potential manufacturing imperfections. The mass of the passive vibrobot is $M=\unit[1.6]{g}$ while its moment of inertia is $J=\unit[232]{g\,mm^2}$.\par
All particles were manufactured from a proprietary photopolymer using a stereolithographic printer.\par

\subparagraph{Features of the experimental setup.}
The motion of active and passive vibrobots occurs on a circular acrylic baseplate with a diameter of \unit[300]{mm} and a height of \unit[15]{mm}, surrounded by a plastic perimeter wall to prevent their escape. The plate tilt is set horizontally with respect to the ground with an accuracy of $\unit[10^{-3}]{degrees}$. This structure is attached to an electromagnetic shaker that induces vibrations of the plate with a frequency of $f$ and vibration amplitude $A=\unit[60(2)]{\upmu m}$ to guarantee stable excitations. 
When not explicitly specified, we set $f=\unit[80]{Hz}$ and $A=\unit[59(2)]{\upmu m}$ (\autoref{Fig:Fig1}, \autoref{Fig:Fig3} and \autoref{Fig:tracerobservables} and \autoref{Fig:appendix}). In contrast, in \autoref{Fig:Fig6}, $f=\unit[80]{Hz}, \unit[100]{Hz}, \unit[120]{Hz}$ according to the legend of the figure. The six points reported are characterized by the values of $f$, $A$, and $E$ listed in \autoref{Tab:freqampresults}.\par
\begin{table}
	\centering
	\begin{tabularx}{0.8\columnwidth}{
		| >{\centering\arraybackslash}X
		| >{\centering\arraybackslash}X
		| >{\centering\arraybackslash}X |
	}
		\hline
		$f$ [Hz] & $A$ [$\upmu$m] & $E$ [mm$^2$ s$^{-2}$]\\
		\hline
			80 & 59 (2) & 442 (1)\\ 
		100 & 47 (2) & 431 (1)\\ 
		100 & 40 (2) & 308 (1)\\ 
		120& 24 (2) & 157 (1)\\
		100& 32 (2)& 204 (1)\\
		120& 28 (2)& 220 (1)\\
		\hline
	\end{tabularx}
	\caption{
		\label{Tab:freqampresults}
		Frequency $f$, amplitude $A$ and corresponding energy $E$ (normalized by the total load). Values for $A$ and $E$ are rounded off to the last significant digit.
	}
\end{table}
The shaker is placed on a concrete block that suppresses resonances. 
In this parameter range, the vibration is uniform across the plate~\cite{Scholz2018inertial}. 
Furthermore the applicable range of $f$ and $A$ has a lower limit, since a minimum acceleration is needed to lift the particles off the plate and an upper limit, where particles start to tumble and fall over. There is no noticeable indication of resonances in the system within the explored frequency range.
Data acquisition occurs by recording images of the system through a high-speed camera with a spatial resolution of 0.3 mm/pixel  and a time resolution of $150$ frames per second.\par

\subsection{Estimate of particle motion parameters}
We track active and passive vibrobots with sub-pixel accuracy, using standard image processing and classical feature recognition methods. Particle positions and orientations in all frames are obtained this way, while translational and angular velocities are approximated by the na\"ive finite difference schemes $\mathbf{V}=(\mathbf{X}(t+\Delta t)-\mathbf{X}(t))/\Delta t$ and $\mathbf{\Omega}=(\mathbf{\Phi}(t+\Delta t)-\mathbf{\Phi}(t))/\Delta t$, respectively, where $\Delta t=\unit[1]{s}/150$. Due to the ﬁnite inertia of the particles, this reflects the particles' instantaneous velocities. In contrast, only the average displacement can be measured in the overdamped regime rendering the velocity a stochastic quantity.\par
Parameters of passive and active particles have been determined by modeling the motion using Langevin equations for a passive and active Brownian particle and fitted to the measurements~\cite{Scholz2018inertial}. Velocity distributions and mean squared displacements are evaluated for an initial set of parameters and the total deviation from the measured results is calculated. This deviation is then minimized using a Nelder-Mead optimization scheme, iteratively improving the model parameters until an optimum is converged upon.\par

\subparagraph{Parameters of an isolated passive tracer.}
The translational dynamics of a passive tracer without active bath ($\phi=0$) is described by a two-dimensional stochastic differential equation
\begin{equation}
	\label{eq:methods_travel_passive}	
	M\dot{\mathbf{V}}+\Gamma_\text{t}\mathbf{V}=\Gamma_\text{t}\sqrt{2D_\text{t}}\boldsymbol\xi\,
\end{equation}
for velocity $\mathbf{V}=\dot{\mathbf{X}}$ and position $\mathbf{X}$,
where $\boldsymbol\xi$ is a white noise term with zero average and unit variance and the coefficients $\Gamma_\text{t}$ and $D_\text{t}$ are the friction and the effective translational diffusion coefficients, respectively. These parameters depend on a complex interplay of factors, such as the intrinsic properties of the particles (mass $M$, shape of the particle and, specifically, of the legs), but also on environmental properties such as the material of the plate where the motion takes place as well as amplitude and frequency of the vibration. By fitting this model to the measurements, we obtain the two free parameters of the translational dynamics $\Gamma_\text{t}=\unit[49.7]{g\,s^{-1}}$ and $D_\text{t}=\unit[1.18]{mm^2\,s^{-1}}$.\par
The rotational dynamics of the passive tracer are given by an inertial equation of motion for the angular velocity $\Omega=\dot\Phi$ where $\Phi$ is the angle of the particle calculated with respect to the $x$-axis
\begin{equation}
	J\dot\Omega=-\Gamma_\text{r}\Omega+\Gamma_\text{r}\sqrt{2D_\text{r}}\xi\,,
\end{equation}
where $\xi$ is a white noise term with zero average and unit variance. The fit yields $\Gamma_\text{r}=\unit[10^4]{g\,mm^2\,s^{-1}}$ and $D_\text{r}=\unit[0.024]{s^{-1}}$, representing the rotational friction and the rotational diffusion coefficients of the passive tracer in the absence of the bath.\par

\subparagraph{Parameters of an isolated active vibrobot.}
The translational dynamics of an active vibrobot is described by a two-dimensional stochastic differential equation
\begin{equation}
\label{eq:methods_travel_active}	m\dot{\mathbf{v}}+\gamma_\text{t}\mathbf{v}=\gamma_\text{t}\sqrt{2D^\text{a}_\text{t}}\boldsymbol\xi^\text{a}+\gamma_\text{t}v_0\mathbf{n}\,
\end{equation}
for velocity $\mathbf{v}=\dot{\mathbf{x}}$ and position $\mathbf{x}$,
where $\boldsymbol\xi^\text{a}$ is a white noise term with zero average and unit variance, and the coefficients $\gamma^\text{a}_\text{t}$ and $D^\text{a}_\text{t}$ are the friction and the effective translational diffusion coefficients, respectively. The term $\gamma v_0\mathbf{n}$ is the active force providing a velocity mode $v_0$ to the particle and an orientation $\mathbf{n}=(\cos\theta,\sin\theta)$, where $\theta$ is the orientational angle of the particle. The fitting method applied to the active particles allows us to estimate $\gamma^\text{a}_\text{t}=\unit[5.9]{g\,s^{-1}}$ and $D^\text{a}_\text{t}=\unit[75]{mm^2\,s^{-1}}$.\par
The rotational dynamics of an active vibrobot are given by an inertial equation for the angular velocity $\omega=\dot{\theta}$ where $\theta$ is the non-wrapping angle of the particle calculated with respect to an arbitrarily chosen null. The angular dynamics reads
\begin{equation}\label{eq:dynamics_omega_active}
	J_\text{a}\dot{\omega}=-\gamma_\text{r}\omega+\gamma_\text{r}\sqrt{2D^\text{a}_\text{r}}\xi^\text{a}\,,
\end{equation}
where $\xi^\text{a}$ is a white noise term with zero average and unit variance. The coefficients $\gamma_\text{r}$ and $D^\text{a}_\text{r}$ represent the rotational friction and the rotational diffusion coefficient of a single active vibrobot, respectively. The numeric fit yields $\gamma_\text{r}=\unit[200]{g\,mm^2\,s^{-1}}$ and $D^\text{a}_\text{r}=\unit[2]{s^{-1}}$.\par
For both passive and active vibrobots, we assume a Stokes friction rather than a Coulomb friction in the dynamics for the translational velocity (\autoref{eq:methods_travel_passive} and \autoref{eq:methods_travel_active}).
Indeed, after kicking a vibrobot, its velocity decays exponentially rather than linearly, as shown in a similar setup~\cite{Scholz2018}.
This originates from the vertical bouncing motion of the particles: The vibrobots' legs are only in contact with the surface briefly to dissipate energy, such that the particle bounces vertically multiple times while the relaxation occurs.

\subsection{Difference between AOUP and ABP dynamics}

Inertial AOUP and inertial ABP dynamics are characterized by equal second moments of the distribution.
Therefore, kinetic temperatures, velocity autocorrelations, and mean-square displacements coincide between the two models~\cite{caprini2022parental, sprenger2023dynamics}.
Differences appear in higher moments and, in general, in the whole shape of the distributions.
Specifically, inertial ABP is characterized by a double-peaked velocity distribution since the active force has a constant modulus.
As a consequence, this model is suitable to describe the dynamics of active vibrobots~\cite{Scholz2018inertial}.
By contrast, the inertial AOUP is characterized by a Gaussian velocity distribution centered in the origin. 
Thus, is the more suitable model to describe the effective dynamics revealed by a passive granular particle immersed in a non-equilibrium bath consisting of active vibrobots.

\subsection{Microscopic derivation of effective tracer dynamics}
To derive a model for the translational dynamics of the tracer we assume that $n_{\text{c}}\ll N$ active vibrobots particles collide with the tracer, where $n_{\text{c}}$ is the number of collisions occurring on the tracer in a typical time window, and $N$ is the total number of bath particles in the system.
Collisions are modeled through elastic forces, such that the dynamics of tracer and active particles will be subject to additional forces $\mathbf{F}$ and $\mathbf{f}_i$, respectively, given by
\begin{flalign}
	\mathbf{F}&=-k\sum_{i=0}^{n_{\text{c}}}\left(\mathbf{X}-\mathbf{x}_i\right)\,,\\
	\mathbf{f}_i&=-k\left(\mathbf{x}_i-\mathbf{X}\right)\,,
\end{flalign}
where $k$ is the parameter of the interaction quantifying the collision strength.\par
In this way, the translational dynamics of passive and active particles are coupled and read
\begin{flalign}
	M\dot{\mathbf{V}}+\Gamma_\text{t}\mathbf{V}&=\Gamma_\text{t}\sqrt{2D_\text{t}}\boldsymbol\xi+\mathbf{F}\,,\\
	m\dot{\mathbf{v}}_i+\gamma_\text{t}\mathbf{v}_i&=\gamma_\text{t}\sqrt{2D^\text{a}_\text{t}}\boldsymbol\xi^\text{a}+\gamma_\text{t}v_0\mathbf{n}_i+\mathbf{f}_i\,,
\end{flalign}
where $\mathbf{n}_i=(\cos\theta_i,\sin\theta_i)$, and $\theta_i$ evolves by \autoref{eq:dynamics_omega_active}. Since we are only interested in the dynamics of the tracer, the effect of interactions between active particles can be mapped onto an effective swim velocity $v_0\to v_0(\phi)$ that explicitly depends on $\phi$ as in previous studies~\cite{cates2015motility}. This assumption is a good approximation, at least, in the low-density regimes considered here, where we expect that $v_0(\phi)$ is only weakly dependent on $\phi$. As shown in Refs.~\cite{sprenger2021time, caprini2022parental}, 
 %\cite{caprini2022role}
the rotational inertia of the active particles can be mapped onto an effective persistence time.
$\mathbf{n}_i$ is a stochastic vector with exponential autocorrelation
\begin{equation}
	\langle\mathbf{n}(t)\cdot\mathbf{n}(s)\rangle=\exp{\left(-\frac t\tau\right)}\,,
\end{equation}
with correlation time $\tau$ given by~\cite{caprini2022role}
\begin{equation}\label{eq:tau_effective_rotinertia}
	\tau\approx\frac1{D_\text{r}^\text{a}}\left(1+\frac{J_\text{a}}{\gamma_\text{r}}D^a_\text{r}\right)\,.
\end{equation}
This approximation holds because our experimental system has rotational inertia $1/D_\text{r}^\text{a}\gg\frac{J_\text{a}}{\gamma_\text{r}}$. By suppressing the particle index because the interactions have been replaced by $v_0(\phi)$ and using $\tau\approx1/D_\text{r}^\text{a}\gg m/\gamma_\text{t}$, an overdamped approximate solution for $\mathbf{x}(t)$ reads
\begin{equation}
	\mathbf{x}(t)\approx e^{-\lambda t}\left[\mathbf{x}(t_0)+\mathop{\textstyle\int\limits_{t_0}^t}\mathop{\mathrm ds}e^{\lambda s}(\lambda\mathbf{X}(s)+\boldsymbol{\Psi}(s))\right]
\end{equation}
where $\lambda=k/\gamma_\text{t}$ and $\boldsymbol{\Psi}$ is a non-Markovian stochastic vector given by
\begin{equation}
	\boldsymbol{\Psi}=\sqrt{2D_\text{t}^\text{a}}\boldsymbol\xi^\text{a}+v_0\mathbf{n}\,.
\end{equation}
We remark that this method has been previously employed by Solon and Horowitz to study a passive tracer in an overdamped active bath~\cite{solon2022einstein}.
Considering the steady-state $t_0\to-\infty$ and $\mathbf{x}(t_0)=0$, and plugging this solution into the equation for $\mathbf{X}(t)$, we immediately obtain
\begin{flalign}\label{eq:methods_dynamics_approx_genlang}
	&M\dot{\mathbf{V}}+\Gamma_\text{t}{\mathbf{V}}+\Gamma_\text{t}\sqrt{2{D_\text{t}}}\boldsymbol{\xi}\nonumber\\
	&\qquad=n_{\text{c}}k\mathbf{X}+n_{\text{c}}ke^{-\lambda t}\int^t\mathop{\mathrm ds}e^{\lambda s}\left(\lambda\mathbf{x}(s)+\boldsymbol{\Psi}(s)\right)\,.
\end{flalign}
By defining $\boldsymbol{\mathcal{I}}(t)=k\int^t\mathop{\mathrm ds}e^{-\lambda(t-s)}\Psi(s)$ and by integrating \autoref{eq:methods_dynamics_approx_genlang} by parts, we obtain
\begin{equation}\label{eq:GeneralLangevin}
	M\dot{\mathbf{V}}+\Gamma_\text{t}\mathbf{V}+\Gamma_\text{t}\sqrt{2D_\text{t}}\boldsymbol\xi=n_\text{c}\left(\boldsymbol{\mathcal{I}}(t)+[\mathcal{K}*\mathbf{V}](t)\right)\,,
\end{equation}
where $*$ denotes the convolution operation and the memory kernel $\mathcal{K}(s)=ke^{-\lambda s}$. \hyperref[eq:GeneralLangevin]{Equation \ref*{eq:GeneralLangevin}} corresponds to a generalized Langevin equation for a tracer in a bath of active particles. The activity is contained in the non-Markovian noise $\boldsymbol{\mathcal{I}}(t)$ with steady-state autocorrelation
\begin{flalign}\label{eq:methods_expressionI}
	\langle\boldsymbol{\mathcal{I}}(t)\cdot\boldsymbol{\mathcal{I}}(t')\rangle&=D_\text{t}^\text{a}\gamma_\text{t}\mathcal{K}(t-t')+\frac{v_0^2k}{\lambda^2-\alpha^2}e^{-\alpha(t-t')}\nonumber\\
	&\enspace+\left[\left(D_\text{t}^\text{a}\gamma_\text{t}+\frac{v_0^2\alpha\gamma_\text{t}}{\alpha^2-\lambda^2}\right)e^{-\lambda(t-t')}\right]\,,
\end{flalign}
where $\alpha:=\tau^{-1}$. The autocorrelation \eqref{eq:methods_expressionI} takes on a particularly simple form in the limit of large spring constant $\tau k/\gamma_\text{t}\gg1$. This condition holds because collisions occur on the shorter time scale of the dynamics. In this limit, $\mathcal{K}(t)\approx\gamma_\text{t}\delta(t)$, so that the tracer will be affected by the effective friction coefficient
\begin{equation}\label{eq:GammaEFF}
	\Gamma_\text{t}^\text{eff}=\Gamma_\text{t}+n_\text{c}\gamma_\text{t}\approx\Gamma_\text{t}\,,
\end{equation}
where we used that $n_\text{c}\ll1$. Furthermore, the autocorrelation of $\boldsymbol{\mathcal{I}}$ reads
\begin{equation}
	\langle\boldsymbol{\mathcal{I}}(t)\cdot\boldsymbol{\mathcal{I}}(t')\rangle\approx D_\text{t}^\text{a}\gamma_\text{t}^2\delta(t-t')+\left[v_0^2\gamma_\text{t}^2e^{-\frac{t-t'}\tau}\right]\,,
\end{equation}
where we neglected $\mathcal O\left(\gamma_\text{t}/k\tau\right)$.\par
Immediately, we can obtain the following equation of motion for the tracer dynamics:
\begin{equation}\label{eq:methods_effectivedynamics}
	M\dot{\mathbf{V}}+\Gamma_\text{t}^\text{eff}\mathbf{V}+\Gamma_\text{t}\sqrt{2D^\text{eff}_\text{t}}\boldsymbol\xi=V_0\Gamma_\text{t}\boldsymbol\eta\,,
\end{equation}
where the typical speed $V_0$ satisfies the relation
\begin{equation}
	V_0\Gamma_\text{t}=n_\text{c}v_0\gamma_\text{t}\,.
\end{equation}
Thus, $V_0$ depends on the packing fraction $\phi$ through the effective swim velocity of the active bath $v_0=v_0(\phi)$. Finally, the effective diffusion coefficient reads $D_\text{t}^\text{eff}=D_\text{t}+n_\text{c}D_\text{t}^\text{a}\approx D_\text{t}$, because $n_\text{c}\ll1$. Together with \autoref{eq:GammaEFF}, these observations allow us to recover the coarse-grained dynamics \eqref{eq:effectivedynamics_translational} from \autoref{eq:methods_effectivedynamics}.\par

\subsection{Derivation of the active Einstein relations}
\subparagraph{Einstein relation for the translational dynamics.}
The translational dynamics of the passive tracer is described by an underdamped equation of motion \eqref{eq:methods_effectivedynamics} subject to additional red noise. This dynamics is known in the literature of active matter as underdamped active Ornstein Uhlenbeck particles (AOUP) that reads
\begin{flalign}\label{eq:methods_dynamicsAOUPunder}
	M\dot{\mathbf{V}}+\Gamma_\text{t}\mathbf{V}&=\Gamma_\text{t}\sqrt{2D_\text{t}}\boldsymbol\xi+\Gamma_\text{t}V_0\boldsymbol\eta\,,\\
	\tau\dot{\boldsymbol\eta}&=-\boldsymbol\eta+\sqrt{2\tau}\boldsymbol\zeta\,,
\end{flalign}
where $\boldsymbol\zeta$ is a white noise term with zero average and unit variance. The model contains two parameters for the translational dynamics, $V_0$ and $\tau=\tau_2$, which depend on the packing fraction $\phi$ through the speed of the active particles, $v_0\to v_0(\phi)$, and the microscopic details of the system. Within our model, the distribution of the translational velocity $\mathbf{V}$ has a Gaussian profile with zero average, $P(\mathbf{V})\propto\exp\left[-M\mathbf{V}^2/(4\mathcal{K}_\text{t})\right]$ fully described by the translational kinetic energy, $\mathcal{K}_\text{t}=M\langle\mathbf{V}^2\rangle/2$, that reads
\begin{equation}\label{eq:theory_kinetic}
	\mathcal{K}_\text{t}=D_\text{t}\Gamma_\text{t}+V_0^2\frac{\tau\Gamma_\text{t}}{1+\tau\Gamma_\text{t}/M}\,.
\end{equation}
This expression contains the dependence on the active bath's packing fraction $\phi$ through the effective speed $V_0\approx n_\text{c}v_0\gamma_\text{t}/\Gamma_\text{t}$, and $\tau$ induced by collisions. Since $\tau\Gamma_\text{t}/M\gg1$, we can aproximate $\mathcal{K}_\text{t}-D_\text{t}\Gamma_\text{t}\propto MV_0^2\propto n_\text{c}^2$.
Previous studies on granular particles predict that the number of collisions per unit time should scale as $n_\text{c}\propto\sqrt\phi$~\cite{Falcon2006} and, consequently, we find the scaling
\begin{equation}
	\mathcal{K}_\text{t}-D_\text{t}\Gamma_\text{t}\propto\phi
\end{equation}
reported in \autoref{eq:scaling}, in agreement with our experimental results (\autoref{Fig:tracerobservables}~B).\par
The autocorrelation of the velocity, $C_V(t)$, of the dynamics \eqref{eq:methods_dynamicsAOUPunder} is given by the sum of two exponentials, as seen in \autoref{Fig:tracerobservables}~C, given by
\begin{equation}
	C_V(t)=\alpha e^{-t\Gamma_\text{t}/M}+\beta e^{-t/\tau}
\end{equation}
where the coefficient $\alpha$ and $\beta$ are given by
\begin{flalign}
	\alpha&=2D_{\text{t}}\frac{\Gamma_{\text{t}}}{M}+2V_0^2\frac{\Gamma_\text{t}}M\frac\tau{1-\tau^2\Gamma_\text{t}^2/M^2}\,,\\
	\beta&=-2V_0^2\frac{\Gamma_\text{t}^2}{M^2}\frac{\tau^2}{1-\tau^2\Gamma_\text{t}^2/M^2}\,.
\end{flalign}
These expressions are consistent with the experimental finding $\tau_1=M/\Gamma_\text{t}$ and the order of magnitude $\tau_2\approx\tau=1/D_\text{r}^\text{a}$.\par
Applying the Green-Kubo relation~\cite{kubo2012statistical}, the translational diffusion coefficient $D_\text{t}^\text{eff}$ can be calculated analytically and reads
\begin{equation}\label{eq:theory_diffcoeff}
	D_\text{t}^\text{eff}=\int^\infty_0\mathop{\mathrm dt}\langle\mathbf{V}(t)\cdot\mathbf{V}(0)\rangle=\alpha\frac M{\Gamma_\text{t}}+\beta\tau\,.
\end{equation}
As expected, $D_\text{t}^\text{eff}$ is enhanced by the effective speed $V_0$, induced by the collisions of the active bath.
By combining \autoref{eq:theory_diffcoeff} and \autoref{eq:theory_kinetic}, $D_\text{t}^\text{eff}$ can be expressed as a function of $\langle\mathbf{V}^2\rangle$, obtaining analytically an active Einstein relation, namely \autoref{eq:Generalized_Einstein_relation} with $\tau_2=\tau$.\par

\subparagraph{Einstein relation for the rotational dynamics.}
The scenario described in \autoref{Fig:Fig3}~G implies that active particles are also able to transfer rotational energy during tapping collisions. This occurs when active bath particles slide around the tracer (tangential collisions) and is consistent with the occurrence of effective alignment interactions. As a crude approximation, we can assume that such a mechanism produces a torque proportional to $\Omega$ that accelerates the rotation. With this idea, the rotational velocity of the tracer evolves through effective underdamped dynamics
\begin{equation}
	J\dot{\Omega}=-\Omega\Gamma_\text{r}^\text{eff}+\Gamma_\text{r}^\text{eff}\sqrt{2D^\text{eff}_\text{r}}\eta\,,
\end{equation}
with $\Gamma_\text{r}^\text{eff}<\Gamma_\text{r}$. The effective torque due to tangential collisions induces an effective decrease in the rotational friction coefficient of the tracer.\par
This model predicts the steady-state properties of $\Omega$ that agree with experimental findings shown in \autoref{Fig:tracerobservables}: The distribution $P(\Omega)$ has a Gaussian-like profile with zero average, $P(\Omega)\propto\exp\left[-J\Omega^2/(2\mathcal{K}_\text{r})\right]$, fully determined by the rotational kinetic energy $\mathcal{K}_\text{r}$ that reads
\begin{equation}
	\mathcal{K}_\text{r}=J\langle\Omega^2\rangle=\Gamma^\text{eff}_\text{r}D^\text{eff}_\text{r}\,,
\end{equation}
where both $D^\text{eff}_\text{r}$ and $\Gamma^\text{eff}_\text{r}$ depend on $\phi$. In addition, the autocorrelation of the angular velocity is given by a single exponential
\begin{equation}
	C_\Omega(t)=\langle\Omega^2\rangle e^{-t\Gamma^\text{eff}_\text{r}/J}
\end{equation}
so that we can identify $\tau_\Omega=J/\Gamma^\text{eff}_\text{r}$. Since the ratio $\langle\Omega^2\rangle/\tau_{\Omega}=D_\text{r}^\text{eff}(\Gamma^{\text{eff}}_\text{r})^2/J^2$ does not depend on $\phi$, as shown in \autoref{Fig:tracerobservables}j, we can conclude that the effective rotational diffusion coefficient scales as $D_\text{r}^\text{eff}\propto\left(\Gamma^{\text{eff}}_\text{r}\right)^{-2}$.\par
Applying the Green-Kubo relation~\cite{kubo2012statistical}, we can calculate analytically the effective rotational diffusion coefficient $D_\text{r}^\text{eff}$ by integrating $C_\Omega(t)$ over time. By expressing $D_\text{r}^\text{eff}$ as a function of $\langle\Omega^2\rangle$, we obtain the Einstein relation for the rotational dynamics (\autoref{eq:Rot_Einstein_rel}),
\begin{equation}
	D_\text{r}^\text{eff}=\int^\infty_0\mathop{\mathrm dt}\langle\Omega(t)\Omega(0)\rangle=\langle\Omega^2\rangle\tau_\Omega\,,
\end{equation}
which was experimentally verified as seen in \autoref{Fig:tracerobservables}~N.\par

\section{Data availability}
The data that support the plots within this paper and other findings of this study are available from the corresponding author upon request.\par

\section{Code availability}
STL files for the design of the active and passive vibrobots is included as Supplementary Information.\par

\bibliographystyle{naturemag}
\bibliography{AOUPbath}

\begin{thebibliography}{10}
\expandafter\ifx\csname url\endcsname\relax
  \def\url#1{\texttt{#1}}\fi
\expandafter\ifx\csname urlprefix\endcsname\relax\def\urlprefix{URL }\fi
\providecommand{\bibinfo}[2]{#2}
\providecommand{\eprint}[2][]{\url{#2}}

\bibitem{marchetti2013hydrodynamics}
\bibinfo{author}{Marchetti, M.} \emph{et~al.}
\newblock \bibinfo{title}{Hydrodynamics of soft active matter}.
\newblock \emph{\bibinfo{journal}{Rev. Mod. Phys.}}
  \textbf{\bibinfo{volume}{85}}, \bibinfo{pages}{1143--1189}
  (\bibinfo{year}{2013}).

\bibitem{Elgeti2015}
\bibinfo{author}{Elgeti, J.}, \bibinfo{author}{Winkler, R.~G.} \&
  \bibinfo{author}{Gompper, G.}
\newblock \bibinfo{title}{{Physics of microswimmers—single particle motion
  and collective behavior: a review}}.
\newblock \emph{\bibinfo{journal}{Rep. Prog. Phys.}}
  \textbf{\bibinfo{volume}{78}}, \bibinfo{pages}{56601} (\bibinfo{year}{2015}).

\bibitem{bechinger2016active}
\bibinfo{author}{Bechinger, C.} \emph{et~al.}
\newblock \bibinfo{title}{Active particles in complex and crowded
  environments}.
\newblock \emph{\bibinfo{journal}{Rev. Mod. Phys.}}
  \textbf{\bibinfo{volume}{88}}, \bibinfo{pages}{045006}
  (\bibinfo{year}{2016}).

\bibitem{mora2016local}
\bibinfo{author}{Mora, T.} \emph{et~al.}
\newblock \bibinfo{title}{Local equilibrium in bird flocks}.
\newblock \emph{\bibinfo{journal}{Nat. Phys.}} \textbf{\bibinfo{volume}{12}},
  \bibinfo{pages}{1153} (\bibinfo{year}{2016}).

\bibitem{ward2008quorum}
\bibinfo{author}{Ward, A.~J.}, \bibinfo{author}{Sumpter, D.~J.},
  \bibinfo{author}{Couzin, I.~D.}, \bibinfo{author}{Hart, P.~J.} \&
  \bibinfo{author}{Krause, J.}
\newblock \bibinfo{title}{Quorum decision-making facilitates information
  transfer in fish shoals}.
\newblock \emph{\bibinfo{journal}{Proc. Natl. Acad. Sci.}}
  \textbf{\bibinfo{volume}{105}}, \bibinfo{pages}{6948--6953}
  (\bibinfo{year}{2008}).

\bibitem{peruani2012collective}
\bibinfo{author}{Peruani, F.} \emph{et~al.}
\newblock \bibinfo{title}{Collective motion and nonequilibrium cluster
  formation in colonies of gliding bacteria}.
\newblock \emph{\bibinfo{journal}{Phys. Rev. Lett.}}
  \textbf{\bibinfo{volume}{108}}, \bibinfo{pages}{098102}
  (\bibinfo{year}{2012}).

\bibitem{wioland2016ferromagnetic}
\bibinfo{author}{Wioland, H.}, \bibinfo{author}{Woodhouse, F.~G.},
  \bibinfo{author}{Dunkel, J.} \& \bibinfo{author}{Goldstein, R.~E.}
\newblock \bibinfo{title}{Ferromagnetic and antiferromagnetic order in
  bacterial vortex lattices}.
\newblock \emph{\bibinfo{journal}{Nat. Phys.}} \textbf{\bibinfo{volume}{12}},
  \bibinfo{pages}{341--345} (\bibinfo{year}{2016}).

\bibitem{liu2021viscoelastic}
\bibinfo{author}{Liu, S.}, \bibinfo{author}{Shankar, S.},
  \bibinfo{author}{Marchetti, M.~C.} \& \bibinfo{author}{Wu, Y.}
\newblock \bibinfo{title}{Viscoelastic control of spatiotemporal order in
  bacterial active matter}.
\newblock \emph{\bibinfo{journal}{Nat.}} \textbf{\bibinfo{volume}{590}},
  \bibinfo{pages}{80--84} (\bibinfo{year}{2021}).

\bibitem{vasarhelyi2018optimized}
\bibinfo{author}{V{\'a}s{\'a}rhelyi, G.} \emph{et~al.}
\newblock \bibinfo{title}{Optimized flocking of autonomous drones in confined
  environments}.
\newblock \emph{\bibinfo{journal}{Sci. Robot.}} \textbf{\bibinfo{volume}{3}},
  \bibinfo{pages}{eaat3536} (\bibinfo{year}{2018}).

\bibitem{rubenstein2014programmable}
\bibinfo{author}{Rubenstein, M.}, \bibinfo{author}{Cornejo, A.} \&
  \bibinfo{author}{Nagpal, R.}
\newblock \bibinfo{title}{{Programmable self-assembly in a thousand-robot
  swarm}}.
\newblock \emph{\bibinfo{journal}{Science}} \textbf{\bibinfo{volume}{345}},
  \bibinfo{pages}{795--799} (\bibinfo{year}{2014}).

\bibitem{Patterson2017}
\bibinfo{author}{Patterson, G.~A.} \emph{et~al.}
\newblock \bibinfo{title}{{Clogging transition of vibration-driven vehicles
  passing through constrictions}}.
\newblock \emph{\bibinfo{journal}{Phys. Rev. Lett.}}
  \textbf{\bibinfo{volume}{119}}, \bibinfo{pages}{248301}
  (\bibinfo{year}{2017}).

\bibitem{PhysRevX.6.011008}
\bibinfo{author}{Mijalkov, M.}, \bibinfo{author}{McDaniel, A.},
  \bibinfo{author}{Wehr, J.} \& \bibinfo{author}{Volpe, G.}
\newblock \bibinfo{title}{Engineering sensorial delay to control phototaxis and
  emergent collective behaviors}.
\newblock \emph{\bibinfo{journal}{Phys. Rev. X}} \textbf{\bibinfo{volume}{6}},
  \bibinfo{pages}{011008} (\bibinfo{year}{2016}).

\bibitem{aranson2007swirling}
\bibinfo{author}{Aranson, I.~S.}, \bibinfo{author}{Volfson, D.} \&
  \bibinfo{author}{Tsimring, L.~S.}
\newblock \bibinfo{title}{Swirling motion in a system of vibrated elongated
  particles}.
\newblock \emph{\bibinfo{journal}{Phys. Rev. E}} \textbf{\bibinfo{volume}{75}},
  \bibinfo{pages}{051301} (\bibinfo{year}{2007}).

\bibitem{kumar2014flocking}
\bibinfo{author}{Kumar, N.}, \bibinfo{author}{Soni, H.},
  \bibinfo{author}{Ramaswamy, S.} \& \bibinfo{author}{Sood, A.}
\newblock \bibinfo{title}{Flocking at a distance in active granular matter}.
\newblock \emph{\bibinfo{journal}{Nat. Commun.}} \textbf{\bibinfo{volume}{5}},
  \bibinfo{pages}{4688} (\bibinfo{year}{2014}).

\bibitem{kudrolli2010concentration}
\bibinfo{author}{Kudrolli, A.}
\newblock \bibinfo{title}{Concentration dependent diffusion of self-propelled
  rods}.
\newblock \emph{\bibinfo{journal}{Phys. Rev. Lett.}}
  \textbf{\bibinfo{volume}{104}}, \bibinfo{pages}{088001}
  (\bibinfo{year}{2010}).

\bibitem{baconnier2022selective}
\bibinfo{author}{Baconnier, P.} \emph{et~al.}
\newblock \bibinfo{title}{Selective and collective actuation in active solids}.
\newblock \emph{\bibinfo{journal}{Nat. Phys.}} \textbf{\bibinfo{volume}{18}},
  \bibinfo{pages}{1234--1239} (\bibinfo{year}{2022}).

\bibitem{VanZuiden2016}
\bibinfo{author}{van Zuiden, B.~C.}, \bibinfo{author}{Paulose, J.},
  \bibinfo{author}{Irvine, W. T.~M.}, \bibinfo{author}{Bartolo, D.} \&
  \bibinfo{author}{Vitelli, V.}
\newblock \bibinfo{title}{{Spatiotemporal order and emergent edge currents in
  active spinner materials}}.
\newblock \emph{\bibinfo{journal}{Proc. Natl. Acad. Sci. U.S.A.}}
  \textbf{\bibinfo{volume}{113}}, \bibinfo{pages}{12919--12924}
  (\bibinfo{year}{2016}).

\bibitem{scholz2021surfactants}
\bibinfo{author}{Scholz, C.}, \bibinfo{author}{Ldov, A.},
  \bibinfo{author}{P{\"o}schel, T.}, \bibinfo{author}{Engel, M.} \&
  \bibinfo{author}{L{\"o}wen, H.}
\newblock \bibinfo{title}{Surfactants and rotelles in active chiral fluids}.
\newblock \emph{\bibinfo{journal}{Sci. Adv.}} \textbf{\bibinfo{volume}{7}},
  \bibinfo{pages}{eabf8998} (\bibinfo{year}{2021}).

\bibitem{lopez2022chirality}
\bibinfo{author}{L{\'o}pez-Casta{\~n}o, M.~A.}, \bibinfo{author}{Seco, A.~M.},
  \bibinfo{author}{Seco, A.~M.}, \bibinfo{author}{Rodr{\'\i}guez-Rivas, {\'A}.}
  \& \bibinfo{author}{Reyes, F.~V.}
\newblock \bibinfo{title}{Chirality transitions in a system of active flat
  spinners}.
\newblock \emph{\bibinfo{journal}{Phys. Rev. Res.}}
  \textbf{\bibinfo{volume}{4}}, \bibinfo{pages}{033230} (\bibinfo{year}{2022}).

\bibitem{buttinoni2013dynamical}
\bibinfo{author}{Buttinoni, I.} \emph{et~al.}
\newblock \bibinfo{title}{Dynamical clustering and phase separation in
  suspensions of self-propelled colloidal particles}.
\newblock \emph{\bibinfo{journal}{Phys. Rev. Lett.}}
  \textbf{\bibinfo{volume}{110}}, \bibinfo{pages}{238301}
  (\bibinfo{year}{2013}).

\bibitem{bricard2013emergence}
\bibinfo{author}{Bricard, A.}, \bibinfo{author}{Caussin, J.-B.},
  \bibinfo{author}{Desreumaux, N.}, \bibinfo{author}{Dauchot, O.} \&
  \bibinfo{author}{Bartolo, D.}
\newblock \bibinfo{title}{Emergence of macroscopic directed motion in
  populations of motile colloids}.
\newblock \emph{\bibinfo{journal}{Nat.}} \textbf{\bibinfo{volume}{503}},
  \bibinfo{pages}{95--98} (\bibinfo{year}{2013}).

\bibitem{fodor2021irreversibility}
\bibinfo{author}{Fodor, {\'E}.}, \bibinfo{author}{Jack, R.~L.} \&
  \bibinfo{author}{Cates, M.~E.}
\newblock \bibinfo{title}{Irreversibility and biased ensembles in active
  matter: Insights from stochastic thermodynamics}.
\newblock \emph{\bibinfo{journal}{Annu. Rev. Condens. Matter Phys.}}
  \textbf{\bibinfo{volume}{13}} (\bibinfo{year}{2021}).

\bibitem{o2022time}
\bibinfo{author}{O’Byrne, J.}, \bibinfo{author}{Kafri, Y.},
  \bibinfo{author}{Tailleur, J.} \& \bibinfo{author}{van Wijland, F.}
\newblock \bibinfo{title}{Time irreversibility in active matter, from micro to
  macro}.
\newblock \emph{\bibinfo{journal}{Nat. Rev. Phys.}}
  \textbf{\bibinfo{volume}{4}}, \bibinfo{pages}{167--183}
  (\bibinfo{year}{2022}).

\bibitem{frey2005brownian}
\bibinfo{author}{Frey, E.} \& \bibinfo{author}{Kroy, K.}
\newblock \bibinfo{title}{Brownian motion: a paradigm of soft matter and
  biological physics}.
\newblock \emph{\bibinfo{journal}{Annalen der Physik}}
  \textbf{\bibinfo{volume}{517}}, \bibinfo{pages}{20--50}
  (\bibinfo{year}{2005}).

\bibitem{hanggi2005introduction}
\bibinfo{author}{H{\"a}nggi, P.} \& \bibinfo{author}{Marchesoni, F.}
\newblock \bibinfo{title}{Introduction: 100 years of brownian motion}
  (\bibinfo{year}{2005}).

\bibitem{einstein1905motion}
\bibinfo{author}{Einstein, A.} \emph{et~al.}
\newblock \bibinfo{title}{On the motion of small particles suspended in liquids
  at rest required by the molecular-kinetic theory of heat}.
\newblock \emph{\bibinfo{journal}{Annalen Der Physik}}
  \textbf{\bibinfo{volume}{17}}, \bibinfo{pages}{208} (\bibinfo{year}{1905}).

\bibitem{kheifets2014observation}
\bibinfo{author}{Kheifets, S.}, \bibinfo{author}{Simha, A.},
  \bibinfo{author}{Melin, K.}, \bibinfo{author}{Li, T.} \&
  \bibinfo{author}{Raizen, M.~G.}
\newblock \bibinfo{title}{Observation of brownian motion in liquids at short
  times: instantaneous velocity and memory loss}.
\newblock \emph{\bibinfo{journal}{Science}} \textbf{\bibinfo{volume}{343}},
  \bibinfo{pages}{1493--1496} (\bibinfo{year}{2014}).

\bibitem{marconi2008fluctuation}
\bibinfo{author}{Marconi, U. M.~B.}, \bibinfo{author}{Puglisi, A.},
  \bibinfo{author}{Rondoni, L.} \& \bibinfo{author}{Vulpiani, A.}
\newblock \bibinfo{title}{Fluctuation--dissipation: response theory in
  statistical physics}.
\newblock \emph{\bibinfo{journal}{Phys. Rep.}} \textbf{\bibinfo{volume}{461}},
  \bibinfo{pages}{111--195} (\bibinfo{year}{2008}).

\bibitem{blickle2007einstein}
\bibinfo{author}{Blickle, V.}, \bibinfo{author}{Speck, T.},
  \bibinfo{author}{Lutz, C.}, \bibinfo{author}{Seifert, U.} \&
  \bibinfo{author}{Bechinger, C.}
\newblock \bibinfo{title}{Einstein relation generalized to nonequilibrium}.
\newblock \emph{\bibinfo{journal}{Phys. Rev. Lett.}}
  \textbf{\bibinfo{volume}{98}}, \bibinfo{pages}{210601}
  (\bibinfo{year}{2007}).

\bibitem{d2003observing}
\bibinfo{author}{D'Anna, G.}, \bibinfo{author}{Mayor, P.},
  \bibinfo{author}{Barrat, A.}, \bibinfo{author}{Loreto, V.} \&
  \bibinfo{author}{Nori, F.}
\newblock \bibinfo{title}{Observing brownian motion in vibration-fluidized
  granular matter}.
\newblock \emph{\bibinfo{journal}{Nat.}} \textbf{\bibinfo{volume}{424}},
  \bibinfo{pages}{909--912} (\bibinfo{year}{2003}).

\bibitem{umbanhowar1996localized}
\bibinfo{author}{Umbanhowar, P.~B.}, \bibinfo{author}{Melo, F.} \&
  \bibinfo{author}{Swinney, H.~L.}
\newblock \bibinfo{title}{Localized excitations in a vertically vibrated
  granular layer}.
\newblock \emph{\bibinfo{journal}{Nat.}} \textbf{\bibinfo{volume}{382}},
  \bibinfo{pages}{793--796} (\bibinfo{year}{1996}).

\bibitem{RN5681}
\bibinfo{author}{Jaeger, H.~M.}, \bibinfo{author}{Nagel, S.~R.} \&
  \bibinfo{author}{Behringer, R.~P.}
\newblock \bibinfo{title}{{Granular Solids, Liquids, and Gases}}.
\newblock \emph{\bibinfo{journal}{Rev. Mod. Phys.}}
  \textbf{\bibinfo{volume}{68}}, \bibinfo{pages}{1259--1273}
  (\bibinfo{year}{1996}).

\bibitem{Agrawal2020}
\bibinfo{author}{Agrawal, M.} \& \bibinfo{author}{Glotzer, S.~C.}
\newblock \bibinfo{title}{{Scale-free, programmable design of morphable chain
  loops of kilobots and colloidal motors}}.
\newblock \emph{\bibinfo{journal}{Proc. Natl. Acad. Sci. U.S.A.}}
  \textbf{\bibinfo{volume}{117}}, \bibinfo{pages}{8700--8710}
  (\bibinfo{year}{2020}).

\bibitem{crisanti2003violation}
\bibinfo{author}{Crisanti, A.} \& \bibinfo{author}{Ritort, F.}
\newblock \bibinfo{title}{Violation of the fluctuation--dissipation theorem in
  glassy systems: basic notions and the numerical evidence}.
\newblock \emph{\bibinfo{journal}{J. Phys. A Math. Theor.}}
  \textbf{\bibinfo{volume}{36}}, \bibinfo{pages}{R181} (\bibinfo{year}{2003}).

\bibitem{maggi2017memory}
\bibinfo{author}{Maggi, C.}, \bibinfo{author}{Paoluzzi, M.},
  \bibinfo{author}{Angelani, L.} \& \bibinfo{author}{Di~Leonardo, R.}
\newblock \bibinfo{title}{Memory-less response and violation of the
  fluctuation-dissipation theorem in colloids suspended in an active bath}.
\newblock \emph{\bibinfo{journal}{Sci. Rep.}} \textbf{\bibinfo{volume}{7}},
  \bibinfo{pages}{17588} (\bibinfo{year}{2017}).

\bibitem{di2010bacterial}
\bibinfo{author}{Di~Leonardo, R.} \emph{et~al.}
\newblock \bibinfo{title}{Bacterial ratchet motors}.
\newblock \emph{\bibinfo{journal}{Proc. Natl. Acad. Sci. U.S.A.}}
  \textbf{\bibinfo{volume}{107}}, \bibinfo{pages}{9541--9545}
  (\bibinfo{year}{2010}).

\bibitem{sokolov2010swimming}
\bibinfo{author}{Sokolov, A.}, \bibinfo{author}{Apodaca, M.~M.},
  \bibinfo{author}{Grzybowski, B.~A.} \& \bibinfo{author}{Aranson, I.~S.}
\newblock \bibinfo{title}{Swimming bacteria power microscopic gears}.
\newblock \emph{\bibinfo{journal}{Proc. Natl. Acad. Sci. U.S.A.}}
  \textbf{\bibinfo{volume}{107}}, \bibinfo{pages}{969--974}
  (\bibinfo{year}{2010}).

\bibitem{loos2020irreversibility}
\bibinfo{author}{Loos, S.~A.} \& \bibinfo{author}{Klapp, S.~H.}
\newblock \bibinfo{title}{Irreversibility, heat and information flows induced
  by non-reciprocal interactions}.
\newblock \emph{\bibinfo{journal}{New J. Phys.}} \textbf{\bibinfo{volume}{22}},
  \bibinfo{pages}{123051} (\bibinfo{year}{2020}).

\bibitem{hokmabad2022chemotactic}
\bibinfo{author}{Hokmabad, B.~V.}, \bibinfo{author}{Agudo-Canalejo, J.},
  \bibinfo{author}{Saha, S.}, \bibinfo{author}{Golestanian, R.} \&
  \bibinfo{author}{Maass, C.~C.}
\newblock \bibinfo{title}{Chemotactic self-caging in active emulsions}.
\newblock \emph{\bibinfo{journal}{Proc. Natl. Acad. Sci. U.S.A.}}
  \textbf{\bibinfo{volume}{119}}, \bibinfo{pages}{e2122269119}
  (\bibinfo{year}{2022}).

\bibitem{ginot2022barrier}
\bibinfo{author}{Ginot, F.}, \bibinfo{author}{Caspers, J.},
  \bibinfo{author}{Kr{\"u}ger, M.} \& \bibinfo{author}{Bechinger, C.}
\newblock \bibinfo{title}{Barrier crossing in a viscoelastic bath}.
\newblock \emph{\bibinfo{journal}{Phys. Rev. Lett.}}
  \textbf{\bibinfo{volume}{128}}, \bibinfo{pages}{028001}
  (\bibinfo{year}{2022}).

\bibitem{thuroff2013critical}
\bibinfo{author}{Th{\"u}roff, F.}, \bibinfo{author}{Weber, C.~A.} \&
  \bibinfo{author}{Frey, E.}
\newblock \bibinfo{title}{Critical assessment of the boltzmann approach to
  active systems}.
\newblock \emph{\bibinfo{journal}{Phys. Rev. Lett.}}
  \textbf{\bibinfo{volume}{111}}, \bibinfo{pages}{190601}
  (\bibinfo{year}{2013}).

\bibitem{petrelli2020effective}
\bibinfo{author}{Petrelli, I.}, \bibinfo{author}{Cugliandolo, L.~F.},
  \bibinfo{author}{Gonnella, G.} \& \bibinfo{author}{Suma, A.}
\newblock \bibinfo{title}{Effective temperatures in inhomogeneous passive and
  active bidimensional brownian particle systems}.
\newblock \emph{\bibinfo{journal}{Phys. Rev. E}}
  \textbf{\bibinfo{volume}{102}}, \bibinfo{pages}{012609}
  (\bibinfo{year}{2020}).

\bibitem{flenner2020active}
\bibinfo{author}{Flenner, E.} \& \bibinfo{author}{Szamel, G.}
\newblock \bibinfo{title}{Active matter: Quantifying the departure from
  equilibrium}.
\newblock \emph{\bibinfo{journal}{Phys. Rev. E}}
  \textbf{\bibinfo{volume}{102}}, \bibinfo{pages}{022607}
  (\bibinfo{year}{2020}).

\bibitem{harada2005equality}
\bibinfo{author}{Harada, T.} \& \bibinfo{author}{Sasa, S.-i.}
\newblock \bibinfo{title}{Equality connecting energy dissipation with a
  violation of the fluctuation-response relation}.
\newblock \emph{\bibinfo{journal}{Phys. Rev. Lett.}}
  \textbf{\bibinfo{volume}{95}}, \bibinfo{pages}{130602}
  (\bibinfo{year}{2005}).

\bibitem{verley2011modified}
\bibinfo{author}{Verley, G.}, \bibinfo{author}{Mallick, K.} \&
  \bibinfo{author}{Lacoste, D.}
\newblock \bibinfo{title}{Modified fluctuation-dissipation theorem for
  non-equilibrium steady states and applications to molecular motors}.
\newblock \emph{\bibinfo{journal}{EPL}} \textbf{\bibinfo{volume}{93}},
  \bibinfo{pages}{10002} (\bibinfo{year}{2011}).

\bibitem{dinis2012fluctuation}
\bibinfo{author}{Dinis, L.}, \bibinfo{author}{Martin, P.},
  \bibinfo{author}{Barral, J.}, \bibinfo{author}{Prost, J.} \&
  \bibinfo{author}{Joanny, J.}
\newblock \bibinfo{title}{Fluctuation-response theorem for the active noisy
  oscillator of the hair-cell bundle}.
\newblock \emph{\bibinfo{journal}{Phys. Rev. Lett.}}
  \textbf{\bibinfo{volume}{109}}, \bibinfo{pages}{160602}
  (\bibinfo{year}{2012}).

\bibitem{neri2019integral}
\bibinfo{author}{Neri, I.}, \bibinfo{author}{Rold{\'a}n, {\'E}.},
  \bibinfo{author}{Pigolotti, S.} \& \bibinfo{author}{J{\"u}licher, F.}
\newblock \bibinfo{title}{Integral fluctuation relations for entropy production
  at stopping times}.
\newblock \emph{\bibinfo{journal}{J. Stat. Mech. Theory Exp.}}
  \textbf{\bibinfo{volume}{2019}}, \bibinfo{pages}{104006}
  (\bibinfo{year}{2019}).

\bibitem{caprini2021generalized}
\bibinfo{author}{Caprini, L.}
\newblock \bibinfo{title}{Generalized fluctuation--dissipation relations
  holding in non-equilibrium dynamics}.
\newblock \emph{\bibinfo{journal}{J. Stat. Mech. Theory Exp.}}
  \textbf{\bibinfo{volume}{2021}}, \bibinfo{pages}{063202}
  (\bibinfo{year}{2021}).

\bibitem{solon2022einstein}
\bibinfo{author}{Solon, A.} \& \bibinfo{author}{Horowitz, J.~M.}
\newblock \bibinfo{title}{On the einstein relation between mobility and
  diffusion coefficient in an active bath}.
\newblock \emph{\bibinfo{journal}{J. Phys. A Math. Theor.}}
  \textbf{\bibinfo{volume}{55}}, \bibinfo{pages}{184002}
  (\bibinfo{year}{2022}).

\bibitem{deseigne2010collective}
\bibinfo{author}{Deseigne, J.}, \bibinfo{author}{Dauchot, O.} \&
  \bibinfo{author}{Chat{\'e}, H.}
\newblock \bibinfo{title}{Collective motion of vibrated polar disks}.
\newblock \emph{\bibinfo{journal}{Phys. Rev. Lett.}}
  \textbf{\bibinfo{volume}{105}}, \bibinfo{pages}{098001}
  (\bibinfo{year}{2010}).

\bibitem{Koumakis2016}
\bibinfo{author}{Koumakis, N.}, \bibinfo{author}{Gnoli, A.},
  \bibinfo{author}{Maggi, C.}, \bibinfo{author}{Puglisi, A.} \&
  \bibinfo{author}{Leonardo, R.~D.}
\newblock \bibinfo{title}{{Mechanism of self-propulsion in 3D-printed active
  granular particles}}.
\newblock \emph{\bibinfo{journal}{New J. Phys.}} \textbf{\bibinfo{volume}{18}},
  \bibinfo{pages}{113046} (\bibinfo{year}{2016}).

\bibitem{szamel2014self}
\bibinfo{author}{Szamel, G.}
\newblock \bibinfo{title}{Self-propelled particle in an external potential:
  Existence of an effective temperature}.
\newblock \emph{\bibinfo{journal}{Phys. Rev. E}} \textbf{\bibinfo{volume}{90}},
  \bibinfo{pages}{012111} (\bibinfo{year}{2014}).

\bibitem{maggi2015multidimensional}
\bibinfo{author}{Maggi, C.}, \bibinfo{author}{Marconi, U. M.~B.},
  \bibinfo{author}{Gnan, N.} \& \bibinfo{author}{Di~Leonardo, R.}
\newblock \bibinfo{title}{Multidimensional stationary probability distribution
  for interacting active particles}.
\newblock \emph{\bibinfo{journal}{Sci. Rep.}} \textbf{\bibinfo{volume}{5}},
  \bibinfo{pages}{10742} (\bibinfo{year}{2015}).

\bibitem{fodor2016far}
\bibinfo{author}{Fodor, {\'E}.} \emph{et~al.}
\newblock \bibinfo{title}{How far from equilibrium is active matter?}
\newblock \emph{\bibinfo{journal}{Phys. Rev. Lett.}}
  \textbf{\bibinfo{volume}{117}}, \bibinfo{pages}{038103}
  (\bibinfo{year}{2016}).

\bibitem{caprini2019activityinduced}
\bibinfo{author}{Caprini, L.}, \bibinfo{author}{Marconi, U. M.~B.} \&
  \bibinfo{author}{Puglisi, A.}
\newblock \bibinfo{title}{Activity induced delocalization and freezing in
  self-propelled systems}.
\newblock \emph{\bibinfo{journal}{Sci. Rep.}} \textbf{\bibinfo{volume}{9}},
  \bibinfo{pages}{1386} (\bibinfo{year}{2019}).

\bibitem{PhysRevLett.129.048002}
\bibinfo{author}{Keta, Y.-E.}, \bibinfo{author}{Jack, R.~L.} \&
  \bibinfo{author}{Berthier, L.}
\newblock \bibinfo{title}{Disordered collective motion in dense assemblies of
  persistent particles}.
\newblock \emph{\bibinfo{journal}{Phys. Rev. Lett.}}
  \textbf{\bibinfo{volume}{129}}, \bibinfo{pages}{048002}
  (\bibinfo{year}{2022}).

\bibitem{scholz2016ratcheting}
\bibinfo{author}{Scholz, C.}, \bibinfo{author}{D'Silva, S.} \&
  \bibinfo{author}{P{\"{o}}schel, T.}
\newblock \bibinfo{title}{{Ratcheting and tumbling motion of vibrots}}.
\newblock \emph{\bibinfo{journal}{New J. Phys.}} \textbf{\bibinfo{volume}{18}},
  \bibinfo{pages}{123001} (\bibinfo{year}{2016}).

\bibitem{digregorio2018full}
\bibinfo{author}{Digregorio, P.} \emph{et~al.}
\newblock \bibinfo{title}{Full phase diagram of active brownian disks: From
  melting to motility-induced phase separation}.
\newblock \emph{\bibinfo{journal}{Phys. Rev. Lett.}}
  \textbf{\bibinfo{volume}{121}}, \bibinfo{pages}{098003}
  (\bibinfo{year}{2018}).

\bibitem{shaebani2020computational}
\bibinfo{author}{Shaebani, M.~R.}, \bibinfo{author}{Wysocki, A.},
  \bibinfo{author}{Winkler, R.~G.}, \bibinfo{author}{Gompper, G.} \&
  \bibinfo{author}{Rieger, H.}
\newblock \bibinfo{title}{Computational models for active matter}.
\newblock \emph{\bibinfo{journal}{Nat. Rev. Phys.}}
  \textbf{\bibinfo{volume}{2}}, \bibinfo{pages}{181--199}
  (\bibinfo{year}{2020}).

\bibitem{Scholz2018inertial}
\bibinfo{author}{Scholz, C.}, \bibinfo{author}{Jahanshahi, S.},
  \bibinfo{author}{Ldov, A.} \& \bibinfo{author}{L{\"{o}}wen, H.}
\newblock \bibinfo{title}{{Inertial delay of self-propelled particles}}.
\newblock \emph{\bibinfo{journal}{Nat. Commun.}} \textbf{\bibinfo{volume}{9}},
  \bibinfo{pages}{5156} (\bibinfo{year}{2018}).

\bibitem{fily2014dynamics}
\bibinfo{author}{Fily, Y.}, \bibinfo{author}{Baskaran, A.} \&
  \bibinfo{author}{Hagan, M.~F.}
\newblock \bibinfo{title}{Dynamics of self-propelled particles under strong
  confinement}.
\newblock \emph{\bibinfo{journal}{Soft Matter}} \textbf{\bibinfo{volume}{10}},
  \bibinfo{pages}{5609--5617} (\bibinfo{year}{2014}).

\bibitem{vladescu2014filling}
\bibinfo{author}{Vladescu, I.} \emph{et~al.}
\newblock \bibinfo{title}{Filling an emulsion drop with motile bacteria}.
\newblock \emph{\bibinfo{journal}{Phys. Rev. Lett.}}
  \textbf{\bibinfo{volume}{113}}, \bibinfo{pages}{268101}
  (\bibinfo{year}{2014}).

\bibitem{Lam2015}
\bibinfo{author}{Lam, K.-D. N.~T.}, \bibinfo{author}{Schindler, M.} \&
  \bibinfo{author}{Dauchot, O.}
\newblock \bibinfo{title}{{Self-propelled hard disks: implicit alignment and
  transition to collective motion}}.
\newblock \emph{\bibinfo{journal}{New J. Phys.}} \textbf{\bibinfo{volume}{17}},
  \bibinfo{pages}{113056} (\bibinfo{year}{2015}).

\bibitem{leoni2020surfing}
\bibinfo{author}{Leoni, M.} \emph{et~al.}
\newblock \bibinfo{title}{Surfing and crawling macroscopic active particles
  under strong confinement: Inertial dynamics}.
\newblock \emph{\bibinfo{journal}{Phys. Rev. Research}}
  \textbf{\bibinfo{volume}{2}}, \bibinfo{pages}{043299} (\bibinfo{year}{2020}).

\bibitem{Scholz2018}
\bibinfo{author}{Scholz, C.}, \bibinfo{author}{Engel, M.} \&
  \bibinfo{author}{P{\"{o}}schel, T.}
\newblock \bibinfo{title}{{Rotating robots move collectively and
  self-organize}}.
\newblock \emph{\bibinfo{journal}{Nat. Commun.}} \textbf{\bibinfo{volume}{9}},
  \bibinfo{pages}{931} (\bibinfo{year}{2018}).

\bibitem{brilliantov2007translations}
\bibinfo{author}{Brilliantov, N.~V.}, \bibinfo{author}{P{\"o}schel, T.},
  \bibinfo{author}{Kranz, W.~T.} \& \bibinfo{author}{Zippelius, A.}
\newblock \bibinfo{title}{Translations and rotations are correlated in granular
  gases}.
\newblock \emph{\bibinfo{journal}{Phys. Rev. Lett.}}
  \textbf{\bibinfo{volume}{98}}, \bibinfo{pages}{128001}
  (\bibinfo{year}{2007}).

\bibitem{caprini2021inertial}
\bibinfo{author}{Caprini, L.} \& \bibinfo{author}{Marini Bettolo~Marconi, U.}
\newblock \bibinfo{title}{Inertial self-propelled particles}.
\newblock \emph{\bibinfo{journal}{J. Chem. Phys.}}
  \textbf{\bibinfo{volume}{154}}, \bibinfo{pages}{024902}
  (\bibinfo{year}{2021}).

\bibitem{nguyen2021active}
\bibinfo{author}{Nguyen, G.~P.}, \bibinfo{author}{Wittmann, R.} \&
  \bibinfo{author}{L{\"o}wen, H.}
\newblock \bibinfo{title}{Active {O}rnstein-{U}hlenbeck model for
  self-propelled particles with inertia}.
\newblock \emph{\bibinfo{journal}{Journal of Physics: Condensed Matter}}
  \textbf{\bibinfo{volume}{34}}, \bibinfo{pages}{035101}
  (\bibinfo{year}{2021}).

\bibitem{Scholz2017}
\bibinfo{author}{Scholz, C.} \& \bibinfo{author}{P{\"{o}}schel, T.}
\newblock \bibinfo{title}{{Velocity distribution of a homogeneously driven
  two-dimensional granular gas}}.
\newblock \emph{\bibinfo{journal}{Phys. Rev. Lett.}}
  \textbf{\bibinfo{volume}{118}}, \bibinfo{pages}{198003}
  (\bibinfo{year}{2017}).

\bibitem{yu2020velocity}
\bibinfo{author}{Yu, P.}, \bibinfo{author}{Schr{\"o}ter, M.} \&
  \bibinfo{author}{Sperl, M.}
\newblock \bibinfo{title}{Velocity distribution of a homogeneously cooling
  granular gas}.
\newblock \emph{\bibinfo{journal}{Physical Review Letters}}
  \textbf{\bibinfo{volume}{124}}, \bibinfo{pages}{208007}
  (\bibinfo{year}{2020}).

\bibitem{eshuis2010experimental}
\bibinfo{author}{Eshuis, P.}, \bibinfo{author}{van~der Weele, K.},
  \bibinfo{author}{Lohse, D.} \& \bibinfo{author}{van~der Meer, D.}
\newblock \bibinfo{title}{Experimental realization of a rotational ratchet in a
  granular gas}.
\newblock \emph{\bibinfo{journal}{Phys. Rev. Lett.}}
  \textbf{\bibinfo{volume}{104}}, \bibinfo{pages}{248001}
  (\bibinfo{year}{2010}).

\bibitem{baldovin2022many}
\bibinfo{author}{Baldovin, M.}, \bibinfo{author}{Caprini, L.},
  \bibinfo{author}{Puglisi, A.}, \bibinfo{author}{Sarracino, A.} \&
  \bibinfo{author}{Vulpiani, A.}
\newblock \bibinfo{title}{The many faces of fluctuation-dissipation relations
  out of equilibrium}.
\newblock In \emph{\bibinfo{booktitle}{Nonequilibrium Thermodynamics and
  Fluctuation Kinetics: Modern Trends and Open Questions}},
  \bibinfo{pages}{29--57} (\bibinfo{publisher}{Springer},
  \bibinfo{year}{2022}).

\bibitem{PhysRevLett.84.3017}
\bibinfo{author}{Wu, X.-L.} \& \bibinfo{author}{Libchaber, A.}
\newblock \bibinfo{title}{Particle diffusion in a quasi-two-dimensional
  bacterial bath}.
\newblock \emph{\bibinfo{journal}{Phys. Rev. Lett.}}
  \textbf{\bibinfo{volume}{84}}, \bibinfo{pages}{3017--3020}
  (\bibinfo{year}{2000}).

\bibitem{PhysRevLett.106.018101}
\bibinfo{author}{Wilson, L.~G.} \emph{et~al.}
\newblock \bibinfo{title}{Differential dynamic microscopy of bacterial
  motility}.
\newblock \emph{\bibinfo{journal}{Phys. Rev. Lett.}}
  \textbf{\bibinfo{volume}{106}}, \bibinfo{pages}{018101}
  (\bibinfo{year}{2011}).

\bibitem{PhysRevLett.103.198103}
\bibinfo{author}{Leptos, K.~C.}, \bibinfo{author}{Guasto, J.~S.},
  \bibinfo{author}{Gollub, J.~P.}, \bibinfo{author}{Pesci, A.~I.} \&
  \bibinfo{author}{Goldstein, R.~E.}
\newblock \bibinfo{title}{Dynamics of enhanced tracer diffusion in suspensions
  of swimming eukaryotic microorganisms}.
\newblock \emph{\bibinfo{journal}{Phys. Rev. Lett.}}
  \textbf{\bibinfo{volume}{103}}, \bibinfo{pages}{198103}
  (\bibinfo{year}{2009}).

\bibitem{kurtuldu2011enhancement}
\bibinfo{author}{Kurtuldu, H.}, \bibinfo{author}{Guasto, J.~S.},
  \bibinfo{author}{Johnson, K.~A.} \& \bibinfo{author}{Gollub, J.~P.}
\newblock \bibinfo{title}{Enhancement of biomixing by swimming algal cells in
  two-dimensional films}.
\newblock \emph{\bibinfo{journal}{Proc. Natl. Acad. Sci. U.S.A.}}
  \textbf{\bibinfo{volume}{108}}, \bibinfo{pages}{10391--10395}
  (\bibinfo{year}{2011}).

\bibitem{maggi2014generalized}
\bibinfo{author}{Maggi, C.} \emph{et~al.}
\newblock \bibinfo{title}{Generalized energy equipartition in harmonic
  oscillators driven by active baths}.
\newblock \emph{\bibinfo{journal}{Phys. Rev. Lett.}}
  \textbf{\bibinfo{volume}{113}}, \bibinfo{pages}{238303}
  (\bibinfo{year}{2014}).

\bibitem{PhysRevLett.99.148302}
\bibinfo{author}{Chen, D. T.~N.} \emph{et~al.}
\newblock \bibinfo{title}{Fluctuations and rheology in active bacterial
  suspensions}.
\newblock \emph{\bibinfo{journal}{Phys. Rev. Lett.}}
  \textbf{\bibinfo{volume}{99}}, \bibinfo{pages}{148302}
  (\bibinfo{year}{2007}).

\bibitem{lowen2020inertial}
\bibinfo{author}{L{\"o}wen, H.}
\newblock \bibinfo{title}{Inertial effects of self-propelled particles: From
  active brownian to active langevin motion}.
\newblock \emph{\bibinfo{journal}{J. Chem. Phys.}}
  \textbf{\bibinfo{volume}{152}}, \bibinfo{pages}{040901}
  (\bibinfo{year}{2020}).

\bibitem{caprini2022parental}
\bibinfo{author}{Caprini, L.}, \bibinfo{author}{Sprenger, A.~R.},
  \bibinfo{author}{L{\"o}wen, H.} \& \bibinfo{author}{Wittmann, R.}
\newblock \bibinfo{title}{The parental active model: A unifying stochastic
  description of self-propulsion}.
\newblock \emph{\bibinfo{journal}{J. Chem. Phys.}}
  \textbf{\bibinfo{volume}{156}}, \bibinfo{pages}{071102}
  (\bibinfo{year}{2022}).

\bibitem{sprenger2023dynamics}
\bibinfo{author}{Sprenger, A.~R.}, \bibinfo{author}{Caprini, L.},
  \bibinfo{author}{L{\"o}wen, H.} \& \bibinfo{author}{Wittmann, R.}
\newblock \bibinfo{title}{Dynamics of active particles with translational and
  rotational inertia}.
\newblock \emph{\bibinfo{journal}{J. Condens. Matter Phys.}}
  \textbf{\bibinfo{volume}{35}}, \bibinfo{pages}{305101}
  (\bibinfo{year}{2023}).

\bibitem{cates2015motility}
\bibinfo{author}{Cates, M.~E.} \& \bibinfo{author}{Tailleur, J.}
\newblock \bibinfo{title}{Motility-induced phase separation}.
\newblock \emph{\bibinfo{journal}{Annu. Rev. Condens. Matter Phys.}}
  \textbf{\bibinfo{volume}{6}}, \bibinfo{pages}{219--244}
  (\bibinfo{year}{2015}).

\bibitem{sprenger2021time}
\bibinfo{author}{Sprenger, A.~R.}, \bibinfo{author}{Jahanshahi, S.},
  \bibinfo{author}{Ivlev, A.~V.} \& \bibinfo{author}{L{\"o}wen, H.}
\newblock \bibinfo{title}{Time-dependent inertia of self-propelled particles:
  The langevin rocket}.
\newblock \emph{\bibinfo{journal}{Phys. Rev. E}}
  \textbf{\bibinfo{volume}{103}}, \bibinfo{pages}{042601}
  (\bibinfo{year}{2021}).

\bibitem{caprini2022role}
\bibinfo{author}{Caprini, L.}, \bibinfo{author}{Gupta, R.~K.} \&
  \bibinfo{author}{L{\"o}wen, H.}
\newblock \bibinfo{title}{Role of rotational inertia for collective phenomena
  in active matter}.
\newblock \emph{\bibinfo{journal}{Phys. Chem. Chem. Phys.}}
  \textbf{\bibinfo{volume}{24}}, \bibinfo{pages}{24910--24916}
  (\bibinfo{year}{2022}).

\bibitem{Falcon2006}
\bibinfo{author}{Falcon, E.} \emph{et~al.}
\newblock \bibinfo{title}{Collision statistics in a dilute granular gas
  fluidized by vibrations in low gravity}.
\newblock \emph{\bibinfo{journal}{Europhys. Lett.}}
  \textbf{\bibinfo{volume}{74}}, \bibinfo{pages}{830} (\bibinfo{year}{2006}).

\bibitem{kubo2012statistical}
\bibinfo{author}{Kubo, R.}, \bibinfo{author}{Toda, M.} \&
  \bibinfo{author}{Hashitsume, N.}
\newblock \emph{\bibinfo{title}{Statistical physics II: nonequilibrium
  statistical mechanics}}, vol.~\bibinfo{volume}{31}
  (\bibinfo{publisher}{Springer Science \& Business Media},
  \bibinfo{year}{2012}).

\end{thebibliography}

\section{Acknowledgments} 
LC acknowledges support from the Alexander Von Humboldt foundation and and acknowledges the European Union MSCA-IF fellowship for funding the project CHIAGRAM.
RW and HL acknowledge support by the Deutsche Forschungsgemeinschaft (DFG) through the SPP 2265, under grant numbers WI 5527/1-1 (RW) and LO 418/25-1 (HL).

\section{Author contributions}
RW, HL, and CS proposed the research.
LC, AL, HE, and CS designed and manufactured the experimental setup.
LC and AL carried out the experiments. LC, AL, and CS analyzed the experimental data. 
LC developed the model and derived the theoretical results.
All authors discussed the results and contributed to writing the manuscript.\par

\section{Competing financial interests}
The authors declare no competing financial interests.\par

\end{document}